%% file: main.tex
\documentclass[10pt,conference,letterpaper]{IEEEtran}
\IEEEoverridecommandlockouts

\usepackage{cite}

\usepackage{enumitem}

\usepackage{amsmath,amssymb,amsfonts}
\usepackage{amsthm} 
\allowdisplaybreaks

\newtheorem{theorem}{Theorem}
\newtheorem{lemma}{Lemma}
\newtheorem{corollary}{Corollary}

\usepackage{graphicx} 
\usepackage{subfig}  
\usepackage{float}    
\usepackage{caption}  

\usepackage{algorithm}
\usepackage{algorithmic}

\usepackage{tabularx}   
\usepackage{booktabs}   
\usepackage{amsmath}    

\usepackage{soul}
\usepackage{textcomp}
\usepackage{xcolor}
\usepackage{CJK}
\usepackage{bm}
\usepackage{fancyhdr}
\usepackage{url}
\usepackage{verbatim}
\usepackage{booktabs}

\usepackage{multirow}
\usepackage{array} 

\usepackage{hyperref}

\usepackage{mathabx}
\usepackage{mathrsfs}
\usepackage{mathtools}
\usepackage{relsize}

\def\BibTeX{{\rm B\kern-.05em{\sc i\kern-.025em b}\kern-.08em
		T\kern-.1667em\lower.7ex\hbox{E}\kern-.125emX}}

\IEEEoverridecommandlockouts\IEEEpubid{\makebox[\columnwidth]{ 979-8-3315-4940-4/25/\$31.00~\copyright~2025 IEEE \hfill} \hspace{\columnsep}\makebox[\columnwidth]{ }} 
\begin{document}
	
\title{\emph{Remoe:} Towards Efficient and Low-Cost MoE Inference in Serverless Computing \\

}

\author{
    \IEEEauthorblockN{
        Wentao Liu\IEEEauthorrefmark{1},
        Yuhao Hu\IEEEauthorrefmark{1},
        Ruiting Zhou\IEEEauthorrefmark{1}\thanks{Corresponding author: Ruiting Zhou (email: ruitingzhou@seu.edu.cn).},
        Baochun Li\IEEEauthorrefmark{2},
        Ne Wang\IEEEauthorrefmark{3}
    }
    \IEEEauthorblockA{ 
        \IEEEauthorrefmark{1}School of Computer Science and Engineering, Southeast University, China \\
        \IEEEauthorrefmark{2}Department of Electrical and Computer Engineering, University of Toronto, Canada \\
        \IEEEauthorrefmark{3}Department of Computing, The Hong Kong Polytechnic University, Hong Kong \\
       Email: 
    \IEEEauthorrefmark{1}(liuwentao, yuhaohu, ruitingzhou)@seu.edu.cn, 
     \IEEEauthorrefmark{2}bli@ece.toronto.edu,
    \IEEEauthorrefmark{3}newang@polyu.edu.hk
    }
}

\maketitle
	
\begin{abstract}

Mixture-of-Experts (MoE) has become a dominant architecture in large language models (LLMs) due to its ability to scale model capacity via sparse expert activation. Meanwhile, serverless computing, with its elasticity and pay-per-use billing, is well-suited for deploying MoEs with bursty workloads. However, the large number of experts in MoE models incurs high inference costs due to memory-intensive parameter caching. These costs are difficult to mitigate via simple model partitioning due to input-dependent expert activation.
To address these issues, we propose \textit{Remoe}, a heterogeneous MoE inference system tailored for serverless computing. \textit{Remoe} assigns non-expert modules to GPUs and expert modules to CPUs, and further offloads infrequently activated experts to separate serverless functions to reduce memory overhead and enable parallel execution.
We incorporate three key techniques: (1) a Similar Prompts Searching (\textit{SPS}) algorithm to predict expert activation patterns based on semantic similarity of inputs; (2) a Main Model Pre-allocation (\textit{MMP}) algorithm to ensure service-level objectives (SLOs) via worst-case memory estimation; and (3) a joint memory and replica optimization framework leveraging Lagrangian duality and the Longest Processing Time (\textit{LPT}) algorithm. We implement \textit{Remoe} on Kubernetes and evaluate it across multiple LLM benchmarks. Experimental results show that \textit{Remoe} reduces inference cost by up to 57\% and cold start latency by 47\% compared to state-of-the-art baselines.

\end{abstract}


\input{introduction}
\input{motivation}

\input{model}
\input{design}

\input{evaluation}
\input{related_work}
\input{conclusion}

\input{appendix}


	

\bibliographystyle{IEEEtran}
\bibliography{reference}

\end{document}

%% file: introduction.tex
\section{Introduction}

The rise of large language models (LLMs) has ushered in a new era of deep learning applications, enabling capabilities such as advanced text generation and context-aware understanding \cite{barreto2023generative, ma2023llm}. Among recent LLM architectures, the Mixture-of-Experts (MoE) model has emerged as a promising solution to scale model capacity without proportionally increasing inference computation. The foundational MoE architecture replaces a transformer's standard feed-forward network (FFN) with multiple expert FFNs and a gating network for token-to-expert routing \cite{lin2024moe}. This approach allows for building vastly larger and more capable models, as only a fraction of the model's total parameters (experts) are used for any given inference task. 
Meanwhile, serverless computing has gained traction as a cost-effective deployment paradigm for machine learning (ML) inference \cite{duan2024mopar}, owing to its elasticity, fine-grained billing, and simplified resource management \cite{li2022serverless}. These features make it particularly attractive for LLM inference workloads that exhibit bursty traffic \cite{fu2024serverlessllm}. 

However, the convergence of MoE models and serverless platforms is far from straightforward. \textbf{Pricing for serverless computing is the product of the resources allocated to a function and its execution duration}. While MoE's sparse expert activation is efficient, its vast number of experts introduces unique challenges under the serverless pricing model.

The primary challenge stems from the massive memory requirement of MoE models. Deploying the full model as a single serverless function typically requires loading all experts into memory, even if most are unused. This results in significant memory waste and high costs during inference, especially when expensive GPU memory is involved. To address the high memory occupation of MoE models, expert offloading has been widely studied \cite{yu2025fmoe, xue2025moe, song2024promoe, tang2024hobbit}, where most experts are cached on slower CPU memory, and only the predicted active experts are dynamically transferred to the GPU for inference. Existing offloading methods such as fMoE \cite{yu2025fmoe} and HOBBIT \cite{tang2024hobbit} implement dynamic expert swapping between the CPU and GPU through experts prefetching techniques. These approaches, however, still require a large, continuously provisioned memory pool on the CPU to hold the inactive experts. This persistent memory allocation fails to eliminate cost inefficiencies thus making existing solutions suboptimal for serverless MoE inference.

To mitigate the high memory costs, distributing experts across multiple serverless functions is a natural strategy. Unfortunately, this approach is complicated by the unbalanced and unpredictable nature of expert activation in MoE. In MoE inference, the activated experts depend heavily on the input prompt and vary across requests. Several studies \cite{yu2025fmoe, song2024promoe, gupta2024lynx} have shown that for a single prompt, expert activation frequencies vary significantly, and this specialized pattern is difficult to predict due to the training method of the gating network \cite{abdin2024phi,dai2024deepseekmoe}. It is challenging for serverless platforms to properly pre-allocate resources for these expert functions with unbalanced workloads.
Current prediction methods, such as \cite{song2024promoe} and \cite{tang2024hobbit}, rely on online, token-by-token predictions during inference. Such an approach is incompatible with serverless environments, because attempting to allocate resources dynamically would result in severe cold start overhead.

Furthermore, this distributed approach introduces a fundamental trade-off. On one hand, deploying experts as multiple functions reduces memory usage per function, but incurs considerable latency due to the communication overhead it incurs. On the other hand, grouping experts into fewer, larger functions reduces the communication overhead, but may lead to memory inefficiency if inactive experts are loaded unnecessarily. While prior work \cite{liu2025optimizing} simplified this by treating each expert as an independent function, it is impractical for modern MoEs. For instance, Deepseek-V3 \cite{liu2024deepseek} contains thousands of experts (256 experts across 61 layers), and managing them as individual functions would create prohibitive deployment and management overhead. Consequently, determining an effective way to partition experts that balances cost and latency presents a major challenge in deploying MoE models in serverless environments. 

To address such high costs, we present \textit{Remoe}, a heterogeneous inference system that minimizes inference costs while satisfying service level objectives (SLOs). To our best knowledge, \textit{Remoe} is the first work to systematically tackle cost-efficient MoE inference in a serverless setting. Highlights of our original contributions are as follows:

\begin{itemize}
    \item \textbf{A Heterogeneous MoE Architecture}. We design a novel architecture that places non-expert modules on GPUs and expert modules on CPUs. Experts are further designated as local (co-located with the main model) or remote (deployed as separate serverless functions), significantly reducing the primary model's memory footprint and enabling parallel inference.

    \item \textbf{Expert Prediction and Resource Pre-Allocation}. We introduce a Similar Prompts Searching (\textit{SPS}) algorithm to predict expert activations via a semantic clustering tree, and a Main Model Pre-allocation (\textit{MMP}) algorithm to pre-allocate main model resources to meet performance SLOs with theoretical guarantees.
    
    \item \textbf{Cost-Latency Optimization for Remote Experts}. We formulate the configuration of remote experts as an optimization problem to balance cost and latency. We develop an efficient optimization framework based on the Lagrangian dual method and a Longest Processing Time (\textit{LPT}) algorithm to determine memory specifications and expert replicas, supported by a formal convexity analysis.
    
    \item \textbf{Prototype Implementation and Experiments}. We implement a prototype of \textit{Remoe} on Kubernetes. On multiple LLM datasets, our experiments show that \textit{Remoe} reduces inference costs by up to 57.1\% and significantly shortens cold start times compared to existing approaches.
\end{itemize}

%% file: motivation.tex
\section{Motivation}
\label{sec:motivation}

\begin{figure}[!htp]
    \centering
    \includegraphics[width=0.8\linewidth]{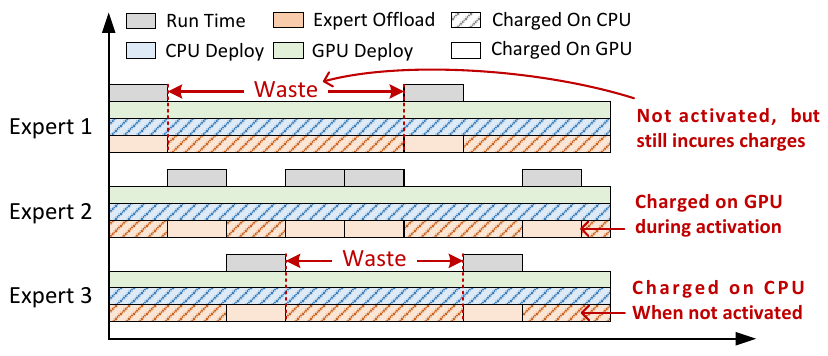}
    \caption{The runtime and charged duration of different deployment methods, and {\em expert offload} represents all expert offloading methods \cite{xue2025moe, tang2024hobbit, song2024promoe, yu2025fmoe} which exchange experts to GPU during activation and offload the remaining experts to CPU.}
    \label{fig:charge_run}
    \vspace{-6mm}
\end{figure}

\textbf{Partial expert activation.} In a serverless context, billing is based on the amount of allocated resources and the execution time. This means that even if most of the experts are not activated, they still occupy memory and incur costs for the entire duration. An example is shown in Fig. \ref{fig:charge_run}. It is clear that whether an MoE model is deployed on a GPU or CPU, all of its experts incur charges for the entire runtime, even if Expert 1 and 3 are each activated just twice. Although expert offloading methods move most of the unused experts to CPUs, all experts still continuously consume memory. A lot of work \cite{tang2024hobbit, xue2025moe, song2024promoe, yu2025fmoe} has shown that the activation frequencies of experts in MoE models differ markedly. To reduce MoE inference cost in a serverless setting, the key is to reduce the memory waste of these low-frequency experts.

\textbf{Communication overhead between layers}.
One bottleneck of Serverless is the limit on the amount of data that can be communicated between functions, also known as the payload size. For example, AWS Lambda has a payload size limit of 6MB for data transmission. To transfer large amounts of data, an intermediary storage service like AWS S3 must be used, which introduces significant latency. For LLM inference, the data transferred between different layers are tokens and their data size is shown in Table \ref{tab:token_size}.

\vspace{-2mm}
\begin{table}[!htp]
\centering
\caption{Token Size for different MoE models (Bfloat16)}
\begin{tabular}{l|ccc}
\hline
\textbf{Model Name}  & \textbf{Parameters} & \textbf{Hidden Size} & \textbf{Token Size}   \\ \hline
Mixtral-8x7B      & 47B        & 4096        & 8 KB   \\ \hline
Mixtral-8x22B     & 141B       & 6144        & 12 KB       \\ \hline
Qwen2-57B-A14B    & 57B        & 3584        & 7 KB \\ \hline
DeepSeek-V2       & 236B       & 5120        & 10 KB       \\ \hline
DeepSeek-V3       & 671B       & 7168        & 14 KB       \\ \hline
Phi-4             & 14.7B      & 5120        & 10 KB       \\ \hline
\end{tabular}
\label{tab:token_size}
\end{table}
\vspace{-3mm}

As we can see, the token size is much smaller than the payload size limit. According to previous work \cite{tang2024hobbit, hwang2024pre}, in low-overhead environments (such as edge computing), requests are often single-batch. Therefore, only a few tokens are transferred between layers during the decoding, which fully meets the payload size limit. This observation makes it feasible to offload low-frequency MoE experts to separate serverless functions (model partitioning) without incurring latency overhead from intermediate storage.


\textbf{Expert inference on the CPU.}
While deploying an entire MoE model on a CPU significantly increases inference latency, its components have varying computational demands. The attention layers are computationally intensive and benefit from GPU acceleration. In contrast, the expert modules are simpler, and since only a few are activated per token, they have lower computational needs. Numerous studies \cite{kvcacheai2024ktransformers, zhong2025hybrimoe} have already validated the feasibility of deploying these experts on CPUs. In a serverless environment where GPUs are much more expensive than CPUs, this enables a cost-saving heterogeneous strategy: run the computationally heavy modules on the GPU and offload the less demanding expert modules to the CPU. Therefore, combining CPU-GPU inference with model partitioning can theoretically reduce the inference cost of MoE models on serverless platforms.

%% file: model.tex
\section{System Model}\label{sec:model}

\subsection{System Overview}\label{sec:system_overview}

In this section, we first consider a general Mixture of Experts (MoE) model. The model is composed of a pre-processing layer $p$, a set of intermediate layers $\mathcal{H} = \{h_1, h_2, \dots, h_L\}$ with length $L$, and a post-processing layer $b$. Each intermediate layer $h_l = (\mathcal{F}_l, \mathcal{E}_l)$ consists of a non-expert module $\mathcal{F}_l$ and an expert module $\mathcal{E}_l$. The non-expert module $\mathcal{F}_l$ is typically composed of transformers and the gate. The expert module is represented as a list $\mathcal{E}_l = \{e_{l,1}, e_{l,2}, \dots, e_{l,K_l}\}$, where $e_{l,k}$ is the $k$-th expert in the $l$-th layer, and $K_l$ is the total number of experts. For certain MoE architectures that share experts, such as DeepseekMoE \cite{dai2024deepseekmoe}, these shared experts are considered part of $\mathcal{F}_l$ since they process all tokens. 

For a request, the inference process of a MoE model can be divided into four stages: \textbf{1) Pre-processing}: The raw natural language is tokenized and encoded by $p$ and the resulting tokens are then passed to $\mathcal{H}$. \textbf{2) Prefilling}: In each layer, all input tokens are processed by $\mathcal{F}_l$ and $\mathcal{E}_l$. The gate routes each token to the most appropriate experts. Finally, the model outputs the most probable token, known as the \textit{first token}. \textbf{3) Decoding}: The \textit{first token} is fed as input to $\mathcal{H}$, and the same computational process is repeated to produce the next token, continuing until all tokens are generated. \textbf{4) Post-processing}: All generated tokens are sent to $b$, converted back into natural language, and then output.

\begin{figure}[!htp]
\vspace{-5mm}
    \centering
    \includegraphics[width=0.9\linewidth]{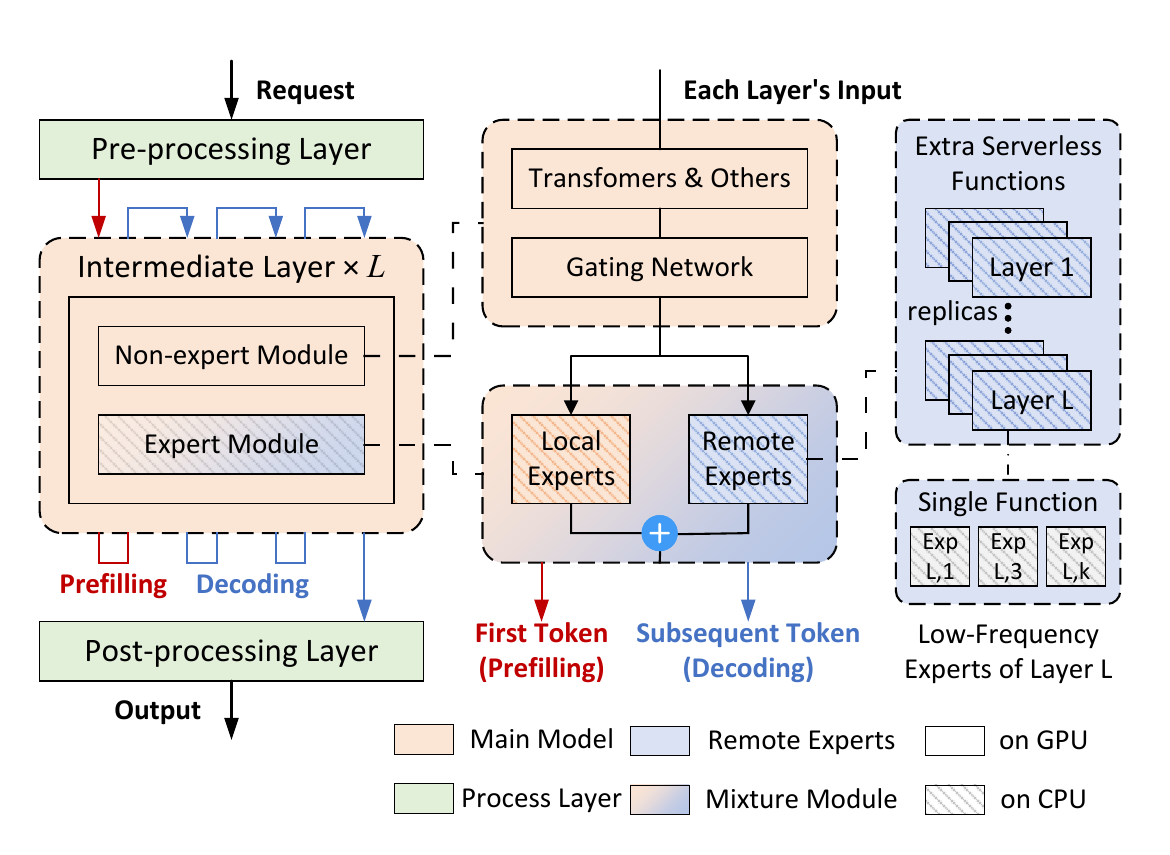}
    \caption{System Overview}
    \label{fig:system_model}
\vspace{-3mm}
\end{figure}

To minimize the inference cost, we design a heterogeneous architecture for \textit{Remoe}. The system overview is shown in Fig. \ref{fig:system_model}. First, we pack all intermediate layers $\mathcal{H}$ as an individual serverless function for inference (main model). The expert module $\mathcal{E}_l$ runs on the CPU; other modules use the GPU. According to the activation frequency, we move some low-frequency experts from the main model to extra serverless functions. For intermediate layer $h_l$, the low-frequency experts in $\mathcal{E}_l$ will be allocated to the same extra function on CPU, and we call them ``remote experts''. The remote expert set of $h_l$ is denoted as $\mathcal{R}_l$. In contrast, those high-frequency experts still remain in the main model, and we call them ``local experts''. This architecture significantly reduces the memory (GPU/CPU) overhead of the runtime container. Meanwhile, the local and remote experts can be computed in parallel, accelerating expert inference.

\textbf{Decision variables}. We introduce four decision variables: 1) Remote expert decision $x_{l,k}$. $x_{l,k}=1$ indicates that the expert $e_{l,k}$ is designated as a remote expert. 2) Remote expert memory $y_{l,v}$. The set of all available memory specifications is denoted by $\mathcal{M} = \{m_1, m_2, \dots, m_V\}$, where $V$ is the total number of specifications. $y_{l,v}=1$ indicates that the memory specification $v$ is allocated to the function holding the remote expert set $\mathcal{R}_l$. 3) Remote expert replicas $z_{l}$. Benefiting from the elastic scaling capabilities of serverless computing, multiple replicas can be instantiated to accelerate the expert inference process. $z_l$ is the replicas number of the functions for $\mathcal{R}_l$. 4) Main model memory $w_v$, $w_v=1$ indicates that the memory specification $v$ is allocated to the main model. On common serverless platforms like AWS Lambda, users only need to set the memory allocation, and the platform automatically assigns corresponding vCPU resources. In this paper, we assume that 1 GB of memory corresponds to 1 vCPU.

\subsection{Inference Latency of \textit{Remoe}}
Since pre-processing and post-processing only involve fixed components and their overhead is typically negligible, we omit these stages and focus on the \textbf{Prefilling} and \textbf{Decoding} phases.

\subsubsection{\textbf{Prefilling}}

The total prefilling time can be presented as:

\vspace{-2mm}
{\small
\begin{equation}
    PT = \sum_{l=1}^{L} (PT_l^f + PT_l^e)
    \label{eq:total_prefill}
\end{equation}
}
\vspace{-3mm}

\noindent where $PT_l^f = \tau_l^f(N^{in})$ is the prefilling time of non-expert module $\mathcal{F}_l$. Here, $\tau_l^f(n)$ is the time for $\mathcal{F}_l$ to process $n$ tokens and $N^{in}$ is the number of input tokens. $PT_l^e$ is the prefilling time of the expert module $\mathcal{E}_l$, which can be expressed as:

\vspace{-4mm}
{\small
\begin{equation}
    PT_l^e = \max [\sum_{k=1}^{K_l} (1-x_{l,k})PT_{l,k}^{loc}, \max_{j \le z_l}\{ZT_{l,j}\} ] + 2\tau^{sw}(N^{in})
    \label{eq:prefill_expert}
\end{equation}
}
\vspace{-4mm}

\noindent where $\sum_{k=1}^{K_l} (1-x_{l,k})PT_{l,k}^{loc}$ and $\max_{j \le z_l}\{ZT_{l,j}\}$ are the end-to-end latency of local experts and remote experts, respectively. We will describe these two parts later. Since expert modules are deployed on CPU, the data need to transfer between GPUs and CPUs twice. $\tau^{sw}(N^{in})$ is used to denote the migration time of $N^{in}$ tokens.

\textbf{Local Experts Latency}. $PT_{l,k}^{loc} = \sum_{v=1}^{V} w_v \tau_{l,k,v}^{c}(N_{l,k}^{pre})$ denotes the prefilling time when $e_{l,k}$ is local, where $w_v$ is the memory specifications allocated to the main model's container, and $\tau_{l,k,v}^{c}(N_{l,k}^{pre})$ represents the computation time for expert $e_{l,k}$ to process $N_{l,k}^{pre}$ tokens under memory specification $v$. The term $N_{l,k}^{pre}$ is the total number of tokens routed to $e_{l,k}$ during prefilling, calculated as the sum $N_{l,k}^{pre} = \sum_{i=1}^{N^{in}} s_{l,k,i}$. $s_{l,k,i} = 1$ indicates the $i$-th input token is processed by $e_{l,k}$.


\textbf{Remote Experts Latency}. With $x_{l,k}$, the remote expert set can be denoted as $\mathcal{R}_l = \{e_{l,k} | x_{l,k} = 1\}$. Since we utilize function replicas to accelerate the remote expert inference, we split the remote expert set $\mathcal{R}_l$ into $\mathcal{R}_{l,1}, \mathcal{R}_{l,2}, \dots, \mathcal{R}_{l,z_l}$ and each replica undertakes the computation of one subset. Different replicas execute simultaneously, so the end-to-end latency of the remote experts is $\max_{j \le z_l}\{ZT_{l,j}\}$. $ZT_{l,j}$ represents the latency for the $j$-th replica. It is calculated as:

\vspace{-3mm}
{\small
\begin{equation}
    ZT_{l,j} = \sum_{e_{l,k} \in \mathcal{R}_{l,j}} ( PT_{l,k}^{rem} + 2 N_{l,k}^{pre}D/B ) + t_{l}^{rem}
    \label{eq:remote_expert_time}
\end{equation}
\vspace{-3mm}
}

\noindent where $PT_{l,k}^{rem} = \sum_{v=1}^{V^e} y_{l,v} \tau_{l,k,v}^{c}(N_{l,k}^{pre})$ is the prefilling time when $e_{l,k}$ is remote. $V^e$ is the total number of memory specifications for remote experts ($V^e < V$). $D$ is the size of a single token embedding and $B$ is the network transfer rate. The term $t_{l}^{rem}$ denotes the additional overhead introduced by the serverless invocation for remote experts of layer $l$ (under warm-start conditions), which is a random variable dependent on the vCPU scheduling policy and resource contention.


\subsubsection{\textbf{Decoding}}

After the first token is generated, the model enters the \textbf{Decoding} stage. Let the total number of generated tokens be $N^{out} + 1$ (including the first token). Decoding time can be expressed as:

\vspace{-5mm}
{\small
\begin{equation}
    GT = \sum_{i=N^{in}+1}^{N^{in}+N^{out}} \sum_{l=1}^{L} (t_l^f + GT_{l,i}^e)
    \label{eq:total_decoding}
\end{equation}
}
\vspace{-3mm}

\noindent where $t_l^f$ is the single token's decoding time of $\mathcal{F}_l$. $GT_{l,i}^e$ is the decoding time of $\mathcal{E}_l$ for token $i$; it can be calculated as:

\vspace{-5mm}
{\small
\begin{equation}
    \begin{split} 
        GT_{l,i}^e =  2\tau^{sw}&(N^{topk}) + \max[\sum_{k=1}^{K_l}(1-x_{l,k})s_{l,k,i} GT_{l,k}^{loc}, \\[-10pt]
        & \quad \sum_{k=1}^{K_l} x_{l,k}s_{l,k,i} (GT_{l,k}^{rem} + 2D/B + t_{l}^{rem})]
    \end{split}
    \label{eq:decoding_expert}
\end{equation}
}
\vspace{-3mm}

\noindent where $N^{topk}$ is the number of experts each token is routed to. $GT_{l,k}^{loc}$ and $GT_{l,k}^{rem}$ are the decoding times of $e_{l,k}$ when it is local and remote, respectively. The former is denoted as $GT_{l,k}^{loc} = \sum_{v=1}^{V} w_v t_{l,k,v}^c$ where $t_{l,k,v}^c$ is the time for expert $e_{l,k}$ to process a single token under memory specification $v$. Similarly, $GT_{l,k}^{rem}$ is calculated as $GT_{l,k}^{rem} = \sum_{v=1}^{V^e} y_{l,v} t_{l,k,v}^c$.





\subsubsection{\textbf{TTFT and TPOT}}

For LLMs, SLOs are typically measured by Time-to-First-Token (TTFT) and Time-per-Output-Token (TPOT). In our model, $T^{cold}$ is the cold start time, and TTFT can be expressed as $T^{ttft} = PT + T^{cold}$. Besides, TPOT is denoted as $T^{tpot} = GT/N^{out}$.


\subsection{Inference Cost of \textit{Remoe}}

Consistent with prior work \cite{yu2021gillis}, we mainly consider the cost of memory usage. We divide the total cost into two parts: the main model cost and remote experts cost.





\subsubsection{\textbf{Main Model Cost}}
The cost of the main model can be calculated as:

\vspace{-5mm}
{\small
\begin{equation}
    C^{loc} = (PT+GT)[c^g M^g + c^e \sum_{v=1}^{V}w_v m_v]
    \label{eq:intermediate_cost}
\end{equation}
}
\vspace{-4mm}

\noindent where $c^c$ is the cost of using 1MB of CPU memory for 1 second, and $M^g$ is the total GPU memory occupied by the main model, which can be expressed as:

\vspace{-3mm}
{\small
\begin{equation}
    M^g = (N^{in} + N^{out})(D+\sum_{l=1}^L a_l) + \sum_{l=1}^L \mu(f_l)
    \label{eq:gpu_memory}
\end{equation}
}
\vspace{-4mm}

\noindent where $a_l$ is the data size of the kv-cache for a single token in layer $l$. Kv-cache technique \cite{vaswani2017attention} prevents the model from re-computing transformer matrices for previous tokens. Consequently, the term $(N^{in} + N^{out})(D + \sum_l a_l)$ represents the total memory occupied by the token embeddings and the entire kv-cache, while $\sum_l \mu(f_l)$ is the memory occupied by the non-expert modules.

\subsubsection{\textbf{Remote Experts Cost}}

The cost associated with the remote experts can be divided into prefilling cost $PC^{rem}$, and the decoding cost $GC^{rem}$. Therefore, the total cost for the remote experts is expressed as $C^{rem} = PC^{rem} + GC^{rem}$.

\textbf{Prefilling Cost}. The cost of remote experts during prefilling is calculated as follows:

\vspace{-5mm}
{\small
\begin{equation}
    PC^{rem} = c^c \sum_{l=1}^{L} \sum_{v=1}^{V^e} y_{l,v} m_v \sum_{j=1}^{z_l} ZT_{l,j}
    \label{eq:prefill_cost_remote}
\end{equation}
}
\vspace{-3mm}

\noindent where the cost of each replica is the product of its memory $y_{l,v}m_v$ and runtime $ZT_{i,j}$.

\textbf{Decoding Cost}. The decoding cost of remote experts, $GC^{rem}$, is calculated as:



\vspace{-4mm}
{\small
\begin{equation}
    \begin{split} 
        GC^{rem} = c^c \sum_{i=N^{in}+1}^{N^{in}+N^{out}} \sum_{l=1}^{L}  \sum_{v=1}^{V^e} y_{l,v}m_v \sum_{k=1}^{K_l}x_{l,k}s_{l,k,i} 
        \\ \cdot (GT_{l,k}^{rem} + 2D/B + t_{l}^{rem})
    \end{split}
    \label{eq:decoding_cost_remote}
\end{equation}
}
\vspace{-3mm}

\noindent where the cost is also the product of its memory $y_{l,v}m_v$ and runtime $(GT_{l,k}^{rem} + 2D/B + t_{l}^{rem})$.

\subsection{Problem Formulation}
The objective is to minimize the total model cost while satisfying SLOs, which is defined as follows:


\vspace{-5mm}
{\small
\begin{subequations}
\label{eq:optimization_problem}
\begin{align}
    \min_{x, y, z, w}&  \quad C^{loc} + C^{rem} \label{eq:objective} \\
    \text{s.t.} \quad & T^{ttft} \le TTFT, \label{eq:constraint_ttft} \\
    & T^{tpot} \le TPOT, \label{eq:constraint_tpot} \\
    & \textstyle\sum_{v=1}^{V^e} y_{l,v} = 1, \label{eq:constraint_y_sum}\\
    & \textstyle\sum_{k=1}^{K_l} x_{l,k}(\mu(e_{l,k}) + DN_{l,k}^{pre}) \le \textstyle\sum_{v=1}^{V^e} y_{l,v} m_v, \label{eq:constraint_remote_mem}\\
     \textstyle\sum_{l=1}^{L} &\textstyle\sum_{k=1}^{K_l} (1-x_{l,k})\mu(e_{l,k}) + DN^{out} \le \textstyle\sum_{v=1}^{V} w_v m_v, \label{eq:constraint_local_mem}\\
    & \textstyle\sum_{e_{l,k} \in \mathcal{R}_{l,j}} N_{l,k}^{pre}D \le U^{payload}, \label{eq:constraint_payload}\\
    & x_{l,k}, y_{l,v}, w_v \in \{0,1\}, \quad \forall l, k, v, \label{eq:constraint_binary}\\
    & z_l \le z^{max}, z_l \in \mathbb{Z}^+, \quad \forall l. \label{eq:constraint_replicas}
\end{align}
\end{subequations}
}
\vspace{-5mm}

Thereinto, Constraint \eqref{eq:constraint_ttft} and \eqref{eq:constraint_tpot} guarantee the TTFT and TPOT. Constraint \eqref{eq:constraint_y_sum} ensures that the remote experts at each layer can only be assigned a single memory specification. Constraint \eqref{eq:constraint_remote_mem} ensures that the allocated memory for remote experts at each layer is sufficient to hold both the model weights and the data for the tokens they process. Similarly, Constraint \eqref{eq:constraint_local_mem} ensures that the memory allocated to the main model is sufficient for its weights and all tokens. Constraint \eqref{eq:constraint_payload} guarantees that the data transferred to a single replica does not exceed the payload size, $U^{payload}$. Finally, Constraints \eqref{eq:constraint_binary} and \eqref{eq:constraint_replicas} define the domains of the decision variables, ensuring the number of expert replicas does not exceed a maximum limit, $z^{max}$.


\textit{\textbf{Challenge}}. In the model described above, unpredictable tokens and complex solutions are two key challenges. In fact, the variable $s_{l,k,i}$ is unknown a priori. Even if all token routing paths were known, the optimization objective remains difficult to solve since it involves products of the decision variables. The situation places the original problem in the category of Nonlinear Programming, which is known to be NP-hard \cite{helmberg1998solving}.

%% file: design.tex
\section{Remoe Design}
\label{sec:design}

\subsection{Main Idea}

To address the challenges previously discussed, we design a system for the MoE inference in serverless, named \textit{Remoe}. When a request arrives, \textit{Remoe} executes the following steps: 

\textit{i. Activation Prediction}. The arriving request is first processed by the pre-processing layer and \textit{Remoe} gets the input tokens. Then \textit{Remoe} invokes \textit{SPS} algorithm to predict the expert activation matrix for the new request. In the offline phase, \textit{Remoe} builds a multi-fork clustering tree based on historical data. Soft Cosine Similarity (SCS) is used to measure the semantic similarity between prompts and build the tree.


\textit{ii. Resource Pre-allocation}. Upon request arrival, \textit{Remoe} employs the \textit{MMP} algorithm to pre-allocate resources. To satisfy TTFT and TPOT constraints, \textit{MMP} determines the optimal remote expert ratio $b$ by estimating the worst-case remote load—a process justified by a proven upper bound. Based on this ratio, it assigns the memory allocation $w_v$ and initiates the main model's cold start.

\textit{iii. Remote experts Selection}. Afterwards, \textit{Remoe} will calculate the expected utility of all experts based on the predicted matrix and set all low-utility experts as remote.

\textit{iv. Memory Optimization}. To reduce the complexity, we construct a new correlation function for $y_{l,v}$ based on their characteristics and fit it, which reformulates the problem. Then, \textit{Remoe} uses Lagrangian duality to solve the problem and the subsequent convexity analysis proves that the optimal solution can be found within the feasible region.

\textit{v. Multi-replicas Inference}. We formulate the multi-replica inference during prefilling as a Multiway Number Partitioning Problem. \textit{Remoe} employs the \textit{LPT} algorithm to solve it, and the resulting upper bound dictates the necessary number of remote expert replicas $z_l$.




\subsection{Activation Distribution Prediction}

For incoming requests, after the pre-processing layer, \textit{Remoe} predicts subsequent expert activation based on the semantic information of input tokens.

\label{sec:prediction}

\begin{figure}[htbp]
    \vspace{-5mm}
    \centering
    \includegraphics[width=0.7\linewidth]{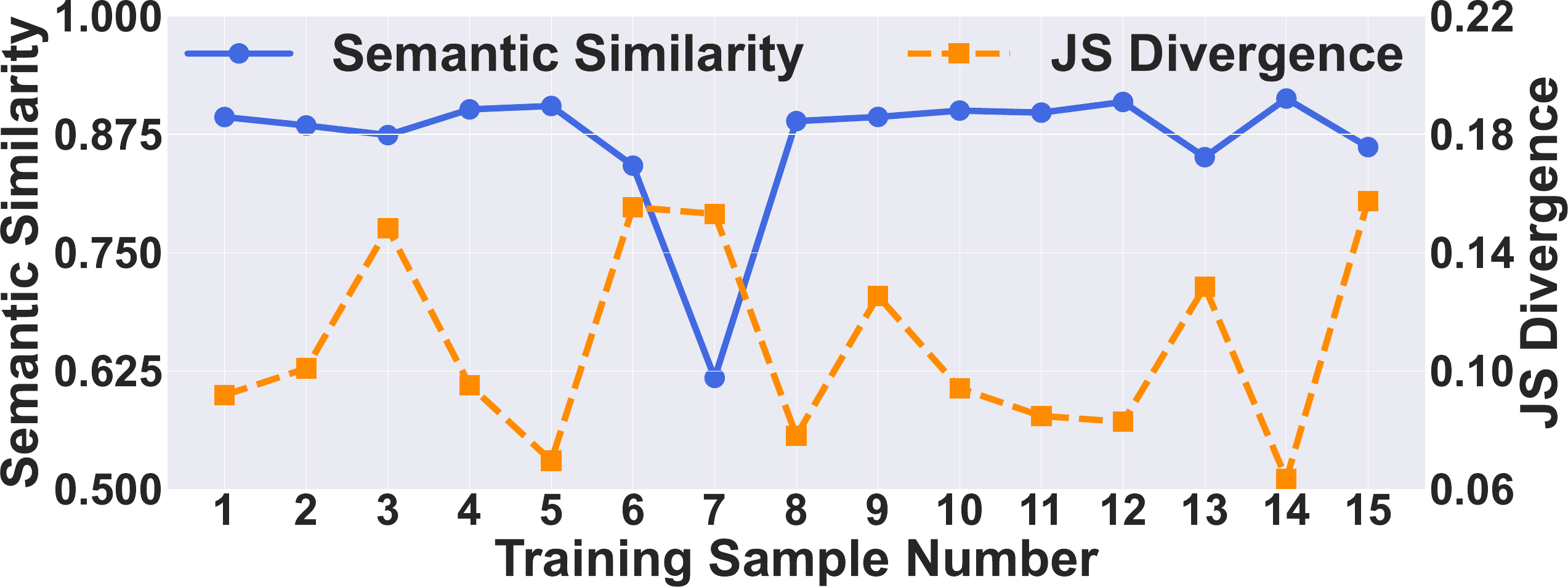}
    \caption{Semantic similarity and expert activation distribution}
    \label{fig:semantic_activation_prove}
\vspace{-4mm}
\end{figure}

Fig. \ref{fig:semantic_activation_prove} compares the semantic similarities and the Jensen-Shannon (JS) Divergence of expert activation distributions between 1 test sample and 15 training samples from LMSYS-Chat-1M dataset \cite{zheng2023lmsys}, fed into GPT2-MoE (Sec. \ref{sec:settings}). Note that JS Divergence is a typical probability distribution similarity comparing method \cite{lee2001effectiveness}. Obviously, semantic similarity positively correlates with expert activation similarity, enabling its use as a proxy for expert activation comparison. Prompt-level expert activation prediction is detailed below.

\textbf{Semantic Similarity Comparison.} We compute the semantic similarity between two prompts, $\zeta_1$ and $\zeta_2$, using SCS \cite{sitikhu2019comparison}. This involves normalizing and concatenating their token embedding matrices, then multiplying by the transpose to yield a symmetric token similarity matrix $\mathbb{C}_{\zeta_1,\zeta_2}$. We also construct two alignment vectors, $\mathbb{V}_1$ and $\mathbb{V}_2$, to mark token ownership per prompt via binary indicators (1: belonging; 0: otherwise). $\mathbb{V}_1$ and $\mathbb{V}_2$ are column vectors.
Thus SCS between semantic embeddings of $\zeta_1$ and $\zeta_2$ is calculated below:

\vspace{-3mm}
{\small
\begin{equation}
    \label{soft_cosine}
    SCS_{1,2} = \frac{\mathbb{V}_1^T \mathbb{C}_{\zeta_1, \zeta_2}\mathbb{V}_2}{\sqrt{\mathbb{V}_1^T \mathbb{C}_{\zeta_1, \zeta_2}\mathbb{V}_1} \cdot \sqrt{\mathbb{V}_2^T \mathbb{C}_{\zeta_1, \zeta_2}\mathbb{V}_2}+\sigma},
\end{equation}}
\vspace{-3mm}

\noindent where $\sigma$ is an extremely small value used to prevent division by zero. Because $\mathbb{C}_{\zeta_1, \zeta_2}$ is a Gram matrix, which is positive semi-definite, $\mathbb{V}_j^T \mathbb{C}_{\zeta_1, \zeta_2}\mathbb{V}_j$ is non-negative.

\textbf{Semantically Similar Prompts Searching.} We efficiently search semantically similar prompts for a new one based on the multi-fork clustering tree.\par
Pairwise semantic similarities for all historical prompts are precomputed. During tree construction, any node (cluster) with more than $\beta$ prompts is recursively partitioned. The partition is based on a customized k-medoids clustering algorithm using prompt-level semantic similarity as distance metric, where roulette wheel sampling-based centroid initialization and subcluster-level centroid updating are conducted.\par
We set $\beta > \alpha$ to augment tree retrieval with local brute-force searching. For a new prompt, the tree is traversed to a leaf by successively selecting the semantically closest subcluster centroid. If there are enough prompts in the leaf, top-$\alpha$ semantically similar ones are returned; otherwise, we turn to the leaf's siblings for supplement.\par

\vspace{-2mm}
\begin{algorithm}
\footnotesize
    \begin{algorithmic}[1] 
        \REQUIRE $\alpha$
        \STATE Initialize clustering tree $tree$
        \WHILE{new prompt $prom$ arrives}
            \STATE $\mathbb{PROM} = []$ 
            \STATE Select one leaf node $leaf$ in $tree$
            \STATE Put samples from $leaf$ into $\mathbb{PROM}$
            \IF{$\text{len}(\mathbb{PROM}) < \alpha$}
                \STATE Turn to $leaf$'s siblings and update $leaf$
                \STATE Add samples into $\mathbb{PROM}$ until $\alpha$ samples are obtained
            \ENDIF
            \RETURN $\mathbb{PROM}$
        \ENDWHILE
    \end{algorithmic}
    \caption{Similar Prompts Searching (\textit{SPS})}
    \label{alg:approx_search} 
\end{algorithm}
\vspace{-2mm}

\textit{SPS} algorithm is outlined in Algorithm \ref{alg:approx_search}, where $\mathbb{PROM}$ represents the set of top-$\alpha$ similar historical prompts searched currently. \textit{SPS} initially builds the clustering tree $tree$ (Line 1). For each new prompt, a leaf $leaf$ is identified to retrieve similar prompts (Lines 2-5). If insufficient samples exist in $leaf$, its siblings are turned to (Lines 6-9). After acquiring $\alpha$ historical prompts, the set $\mathbb{PROM}$ is returned (Lines 10-11).

\textbf{Expert Activation Distribution Prediction.} For each historical prompt $\zeta_j$, we obtain its expert activation distribution matrix $\tilde{S}$. Matrix element $\tilde{s}_{l,k}=\frac{frec_{l,k}}{\sum_k{frec_{l,k}}}$ represents the ``linear scaling activation frequency" of expert $e_{l,k}$ during prefilling of $\zeta_j$. $frec_{l,k}$ is the times $e_{l,k}$ is activated. $\sum_k{frec_{l,k}}$ equals product of the number of $\zeta_j$'s tokens $N^{in}_j$ and the number of experts activated by one token in each layer $N^{topk}$.\par
SCS between the new prompt and the retrieved $\alpha$ historical prompts are converted into probability weights via softmax. The expert activation distribution matrices of historical prompts are then weighted-summed to predict the result.

\subsection{Resource Pre-allocation for Main Model}
\label{sec:pre_allocate}

To handle a cold start, \textit{Remoe} pre-allocates memory for the main model as soon as a request arrives. This mechanism is separate from activation prediction (Sec. \ref{sec:prediction}). This parallel approach is effective because the main model's pre-allocation can overlap with the pre-processing layer's cold start, which must complete before activation prediction can begin.

\begin{figure}[!htp]
\vspace{-4mm} 
    \begin{minipage}[b]{0.47\linewidth}
        \centering
        \includegraphics[width=\linewidth]{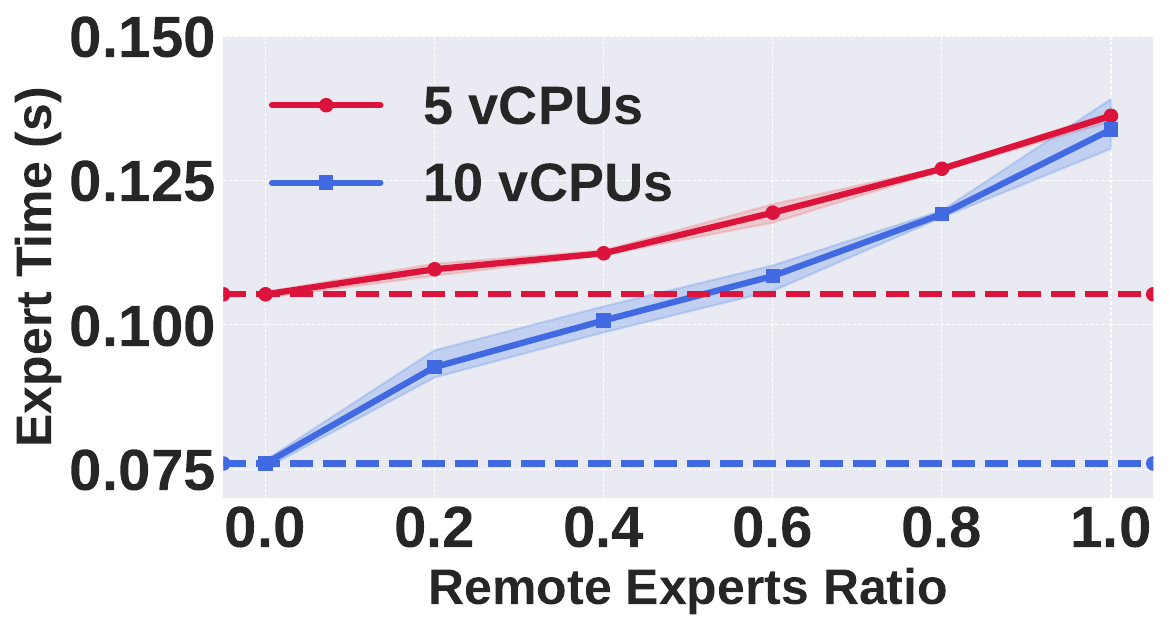}
        \caption{Expert inference time with 5 and 10 cores}
        \label{fig:remote_compare}
    \end{minipage}
    \hspace{0.2mm} 
    \begin{minipage}[b]{0.49\linewidth}
        \centering
        \includegraphics[width=\linewidth]{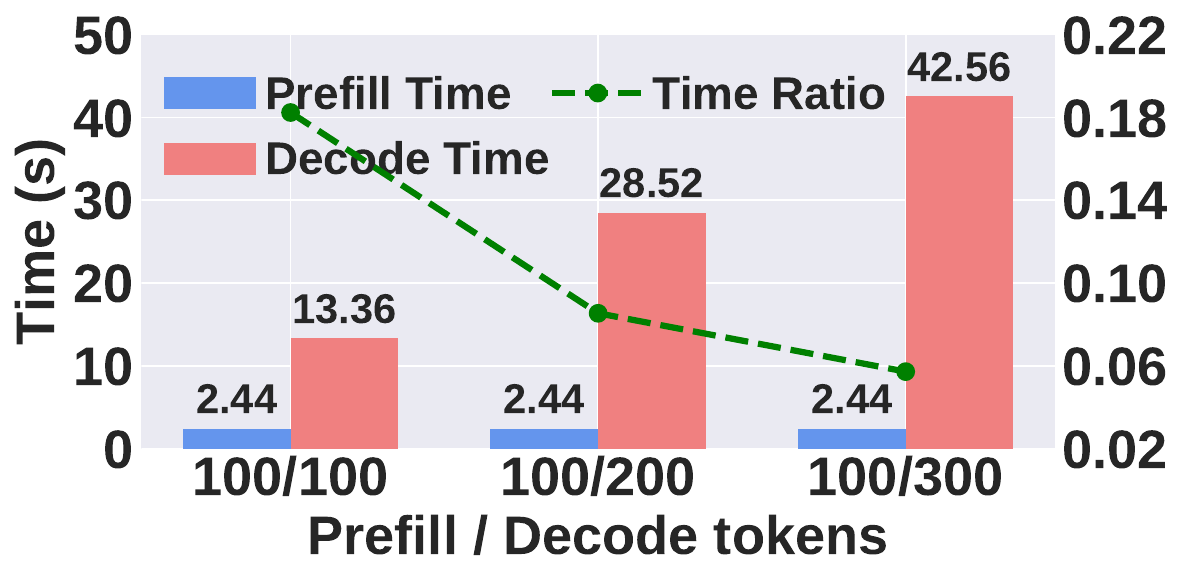}
        \caption{Prefilling Time vs. Decoding Time}
        \label{fig:prefill_decode_compare}
    \end{minipage}
    
\vspace{-5mm} 
\end{figure}

\textbf{Decoding Time Analysis}. We simplify Eq. \eqref{eq:decoding_expert} by removing the max function, assuming the remote expert path is always the performance bottleneck. This assumption is supported by two key observations. First, as shown in Fig. \ref{fig:remote_compare}, the expert inference time increases nearly linearly with the ratio of remote experts. This indicates that, with the same vCPUs, remote experts dominate the inference time in Eq. \eqref{eq:decoding_expert}. Second, in practical scenarios, the main model is typically allocated more vCPUs, ensuring faster computation for local experts.

\begin{theorem}
\label{theo:one_max_tokens}
When $n$ tokens pass through layer $l$, the number of tokens processed by the $k$-th expert will not exceed $\frac{\sqrt{3n}}{2} + \frac{n}{K_l}$ with a high probability (95\%).
\end{theorem}

\begin{corollary}
\label{cor:m_max_tokens}
For $n$ tokens and $m$ experts, processed tokens will not exceed $\frac{\sqrt{3n}}{2} + \frac{mn}{K_l}$ with a high probability (95\%).
\end{corollary}

\textbf{Main Model Pre-allocation}. For the main model, we must pre-allocate a minimum memory specification that guarantees SLOs are met even in the worst-case scenario. Theorem \ref{theo:one_max_tokens} and Corollary \ref{cor:m_max_tokens} provide an upper bound in such a scenario. To this end, we design the Main Model Pre-allocation (\textit{MMP}) algorithm detailed in Algorithm \ref{alg:mmp}.

\vspace{-2mm}
\begin{algorithm}
\footnotesize
\caption{Main Model Pre-allocation (\textit{MMP})}
\label{alg:mmp} 
\begin{algorithmic}[1] 
    \REQUIRE $V^e$;
    \STATE Initialize $M^{min} = \sum_{l=1}^{L} \sum_{k=1}^{K_l} (1-x_{l,k})\mu(e_{l,k}) + N^{max}D$,
    \STATE Initialize remote expert ratio $b\leftarrow1$, $M^{cal} \leftarrow m_{V^e}$
    \REPEAT
        \FOR{$l = 1$ to $L$}
            \STATE Calculate the remote time based on Corollary \ref{cor:m_max_tokens} and $b$
        \ENDFOR
        \STATE Calculate the memory of local experts $M^e$ with $b$
        \STATE Set the main model memory $M \leftarrow \max(M^{min} + M^e, M^{cal})$
        \STATE Calculate the TTFT and TPOT with $M$ and $b$.
        \STATE {$b \leftarrow b - \epsilon$}
    \UNTIL{TTFT and TPOT limits are met}
    \STATE Select the minimum specification $w_v$ that satisfies $m_{w_v} \ge M$
    \RETURN $w_v$
\end{algorithmic}
\end{algorithm}
\vspace{-3mm}

First, \textit{MMP} initializes the minimum memory $M^{min}$ for non-expert modules caching. It also sets remote expert ratio $b$ and $M^{cal}$, the minimum memory required to ensure local experts execute faster than remote ones (Lines 1-2). With specific $b$, \textit{MMP} first calculates the remote processing time of each layer based on Corollary \ref{cor:m_max_tokens} and $b$ (Lines 4-6). This allows for the calculation of the worst-case remote inference latency. Then, \textit{MMP} calculates the memory required to cache local experts for a given ratio $b$ (Line 7). According to it, the main model memory is confirmed and $M^{min} + M^e$ is the minimum memory to hold the parameters (Line 8). This process is repeated with decreasing values of $b$ until both TTFT and TPOT are met (Lines 9-11). Finally, \textit{MMP} returns the minimum specification $w_v$, such that $m_{w_v} \ge M$ (Lines 12-13).


\subsection{Remote Experts Selection}

Given the expert activation matrix $\tilde{S}$ and ratio $b$, we first calculate the expected number of tokens for each $e_{l,k}$. For the prefilling, this is $E[N_{l,k}^{pre}] = N^{in} \tilde{s}_{l,k}$, and for the decoding, it is $E[N_{l,k}^{dec}] = N^{out}N^{topk}\tilde{s}_{l,k}$. Our objective is to minimize latency for a given remote expert ratio, $b$, which is obtained in Sec. \ref{sec:pre_allocate}. To achieve this, we define a utility score $u_{l,k} = E[N_{l,k}^{pre}] + E[N_{l,k}^{dec}]$ and choose the experts with the lowest utility scores to be remote. This selection is formally defined as choosing the set of remote experts, $R_l$, such that: $\mathcal{R}_l = \mathop{\arg\min}_{\mathcal{R}_l} \sum_{e_{l,k} \in \mathcal{R}_l}u_{l,k}, |\mathcal{R}_l|=bK_l, \forall l$. All $x_{l,k}$ are set according to $\mathcal{R}_l$.



\subsection{Remote Experts Memory Optimization}

With $w_v$ and $X_l$, the original problem transforms into an optimization problem of variables $y_{l,v}$ and $z_l$. 

We observe that: \textbf{1) Looser TTFT constraint:} The expert replica decision variable $z_l$ only exists in the prefilling stage. Due to the cold start time $T^{cold}$, the TTFT constraint is often looser than the TPOT constraint. \textbf{2) Longer decoding time:} During the prefilling, each expert layer undergoes batch processing only once. Therefore, this stage is much shorter than decoding with multiple iterations \cite{liu2023deja}. Fig. \ref{fig:prefill_decode_compare} shows the prefilling/decoding times for different numbers of tokens.


\textbf{Problem Reformulation}. Based on these observations, the contribution of variable $y_{l,v}$ to the optimization objective is considered to be concentrated in the decoding stage. Therefore, we can fix the prefilling time as a ratio of the decoding time to serve as an upper bound ($PT \le \eta GT$), and usually $\eta \le 0.1$ according to Fig. \ref{fig:prefill_decode_compare}. After removing all constant values unrelated to $y$, the optimization objective for the memory allocation of remote experts can be expressed as:

\vspace{-5mm}
{\small
\begin{equation}
\label{eq:simplified_problem_v2}
    \min_{y} \quad P_1 = (1+\eta)\sum_{l=1}^L \sum_{v=1}^{V^e} y_{l,v}(\tilde{s}_l T_{l,v}^{rem} + t_l^{rem})(H^w + c^c m_v) 
\end{equation}
}
\vspace{-4mm}


\noindent The remaining constraints are similar to those in Eq. \eqref{eq:optimization_problem} and are omitted here due to space limit. Here, the constant $H^w = c^g M^g + c^c \sum_{v'=1}^V w_{v'} m_{v'}$ is the overhead per unit time of the main model. $T_{l,v}^{rem} = N^{topk} t_{l,v}^c$ is the computation time for remote experts to decode all tokens (number here is $N^{topk}$). $\tilde{s}_l = \sum_{k=1}^{K_l} x_{l,k} \tilde{s}_{l,k}$ is the total probability of each token transferred to those remote experts.

\begin{figure}[htbp]
\vspace{-7mm}
    \centering
    \subfloat[GPT2-moe]{\includegraphics[width=.48\columnwidth]{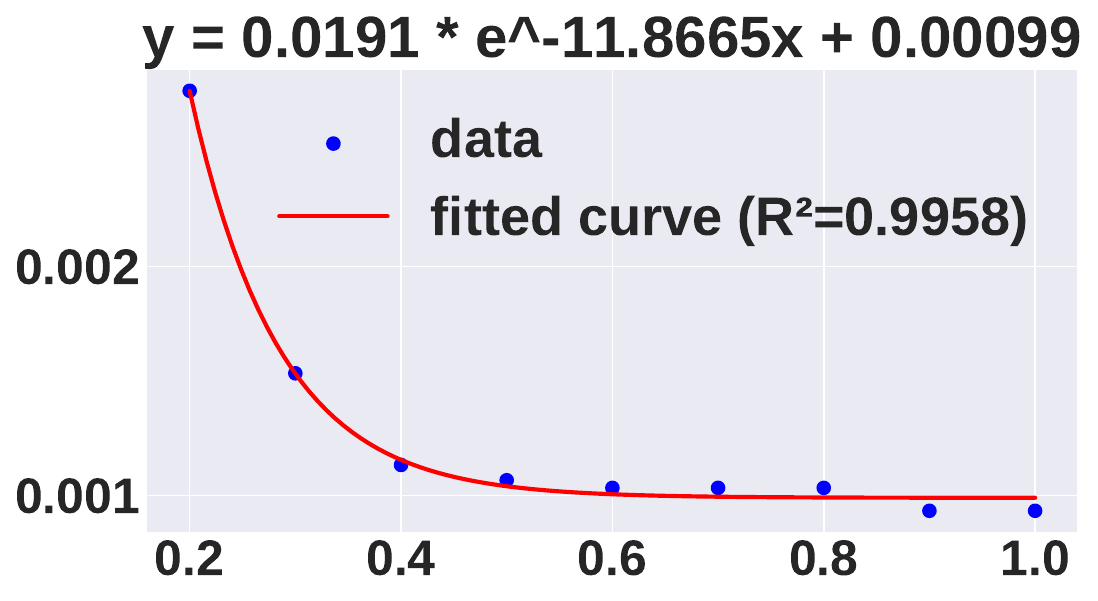}}\hspace{5pt}
    \subfloat[Deepseek-v2-lite]{\includegraphics[width=.48\columnwidth]{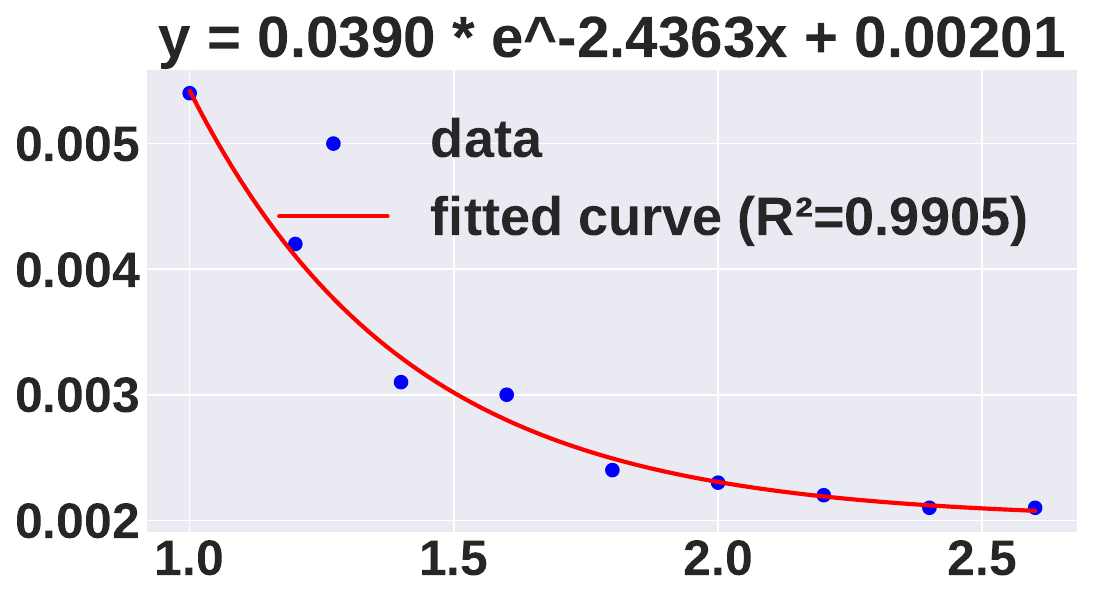}\vspace{-5mm}}\\[-5pt]
    \caption{Fitted Curves of CPU Resources vs. Inference Time}
    \label{fig:fit_curve}
\vspace{-3mm}
\end{figure}

\textbf{Function Construction and Fitting}. For this problem, the search space for memory size is large, and the solution complexity remains high. Therefore, we linearize the discrete term $\sum_{v=1}^{V^e} y_{l,v} m_v$ into a continuous variable $\tilde{y}_l$, where $m_1 \le \tilde{y}_l \le m_{V^e}$. We consider that the inference time of remote experts gradually decreases as the allocated memory increases, eventually converging to a constant. To model this characteristic, we construct the formula $\tilde{T}_l^{rem} = \theta_1 \exp(-\theta_2 \tilde{y}_l) + \theta_3$ ($\theta_1, \theta_2, \theta_3 > 0$). The parameters herein can be obtained by fitting data from model profiling, as illustrated by the fitted curve in Fig. \ref{fig:fit_curve} and the objective can be transformed into $P_2$:


\vspace{-5mm}
{\small
\begin{equation}
\label{eq:final_problem}
    \min_{y} \quad P_2 = (1+\eta)\sum_{l=1}^L \tilde{s}_l\left(\tilde{T}_l^{rem} + \frac{t_l^{rem}}{\tilde{s}_l}\right)\left(H^w + c^c\tilde{y}_l\right) 
\end{equation}
}
\vspace{-3mm}


Although all integer terms have been relaxed into continuous ones, the objective function introduces non-linear terms such as $H^w \tilde{T}_l^{rem}$ and $c^c \tilde{y}_l \tilde{T}_l^{rem}$, making it unsolvable by linear programming.

\textbf{Convexity Analysis}. To enable the subsequent optimization, we first perform the convexity analysis on the constructed functions and objective function.

\begin{theorem}
\label{theo:convex_analysis}
Let $g(\tilde{y}_l) = \left(\tilde{T}_l^{rem} + \frac{t_l^{rem}}{\tilde{s}_l}\right)\left(H^w + c^c\tilde{y}_l\right)$. For $\tilde{y}_l \in [\frac{2}{\theta_2} - \frac{H^w}{c^c}, \infty)$, the function $g(\tilde{y}_l)$ is strictly convex and continuously differentiable. And when $\theta_2 \ge \frac{2c^c}{H^w}$, the function $g(\tilde{y}_l)$ is strictly convex on $(0, \infty)$.   
\end{theorem}


For Theorem \ref{theo:convex_analysis}, we need to analyze whether different models satisfy this characteristic. As shown in Fig. \ref{fig:fit_curve}, the values of $\theta_2$ for GPT2-moe and Deepseek-v2-lite are 11.8665 and 2.4363, respectively. On commercial serverless platforms that support GPU resource allocation (e.g., Alibaba Cloud, Tencent Cloud), the overall cost standard for GPU is generally 3 times or more than that of CPU, i.e., $\frac{c^g}{c^c} \ge 3$. Therefore, we have: $\frac{2c^c}{H^w} = \frac{2}{c^g M^g / c^c+ \sum_{v'=1}^V w_{v'}m_{v'}} \le \frac{2}{3M^g + \sum_{v'=1}^V w_{v'}m_{v'}}$. Here, $M^g$ is the GPU memory overhead of the non-expert layers, and $\sum_{v'=1}^V w_{v'}m_{v'}$ is the CPU memory overhead of the main model. For Deepseek-v2-lite, its non-expert layers have approximately 0.5B parameters, even if only 3GB of memory is allocated to the main model, we have $\frac{2c^c}{H^w} \approx 0.25 \ll 2.4363$. Under a similar analysis, when the main model retains only 12.5\% of the experts as local, the value for GPT2-moe is $\frac{2c^c}{H^w} \approx 2.72 \ll 11.8665$. It can be seen that most MoE models conform to the aforementioned characteristic.


\textbf{Lagrangian Solving}. After analyzing the convexity of problem $P_2$, we give the dual problem of the primal problem $P_2$, denoted as $P_2^D$:

\vspace{-6mm}
{\small
\begin{subequations}
\begin{align}
     \max_\lambda \quad & P_2^D = (1+\eta)\sum_{l=1}^L \tilde{s}_l g(\tilde{y}_l) + \sum_{j=1}^4\sum_{l=1}^L \lambda_{l,j}q_{l,j}^c(\tilde{y}_l) \quad \\[-3pt]
    \text{s.t. } \quad & \lambda_{l,1}, \lambda_{l,2}, \lambda_{l,3}, \lambda_{l,4} \ge 0, \forall l
\end{align}
\end{subequations}
}
\vspace{-5mm}

\noindent where $q^c_{l,j}(\tilde{y}_l)$ represents the j-th constraint function in problem $P_2$, and $\lambda_{l,j}$ is the corresponding dual variable. Thereinto, $q^c_{l,1}(\tilde{y}_l)$ is the TPOT constraint and the rest are linear constraints on the range of $\tilde{y}_l$.


\begin{lemma}[Slater's Condition]
\label{lem:slater}
All constraints $q^c_{l,j}(\tilde{y}_l)$ are convex, and when $g(\tilde{y}_l)$ is strictly convex on $(0, \infty)$, problem $P_2$ is a convex optimization problem and strong duality holds.
\end{lemma}

\begin{theorem}
\label{theo:opt_solu}
Let $\tilde{y}^*; \lambda^*$ be the solution to the dual problem $P_2^D$ that satisfies the KKT conditions. Then $\tilde{y}^*$ is also the optimal solution to the primal problem $P_2$.
\end{theorem}

\noindent According to Theorem \ref{theo:opt_solu}, the problem $P_2$ can be solved using the Lagrangian duality method, and the resulting remote expert memory $y_{l,v}$ is the optimal solution for this problem.

\subsection{Remote Experts Multi-replicas Inference}

\subsubsection{\textbf{Remote Expert Subsets Partitioning}}

\begin{figure}[!htp]
    \vspace{-5mm}
    \centering
    \includegraphics[width=0.7\linewidth]{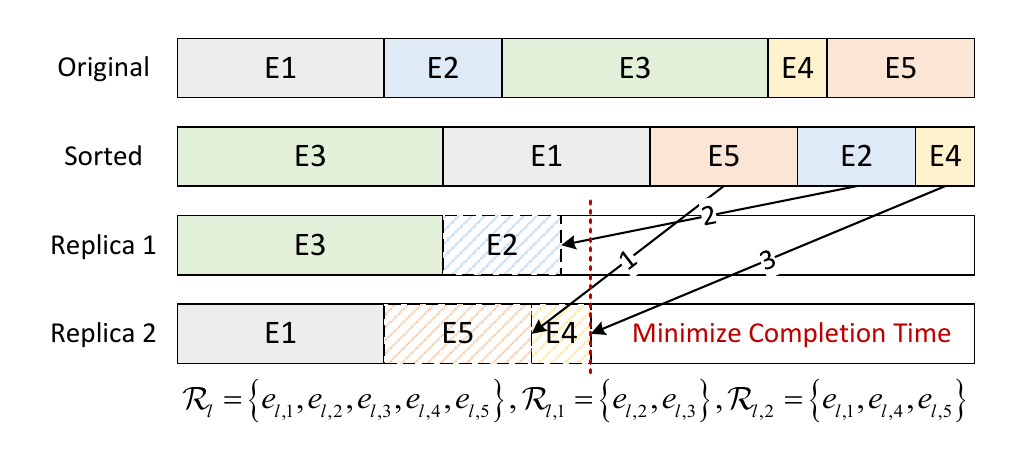}
    \vspace{-3mm}
    \caption{Multiway Number Partitioning problem and \textit{LPT}}
    \label{fig:lpt}
    \vspace{-4mm}
\end{figure}

In Eq. \eqref{eq:remote_expert_time}, we discussed partitioning the set $\mathcal{R}_l$ into $\mathcal{R}_{l,1} \dots, \mathcal{R}_{l,z_l}$. To minimize $\max_{j}\{ZT_{l,j}\}$, we model the optimal partition as a Multiway Number Partitioning problem. An example is in Fig. \ref{fig:lpt}.

Our objective is to assign tasks to different replicas, minimizing the completion time of all replicas. The subset $\mathcal{R}_{l,1}$ correspond to remote expert tasks handled by replica 1. We use \textit{LPT} algorithm to solve it. In simple terms, \textit{LPT} sorts the tasks, and always selects the replica with the minimum load to assign tasks sequentially. The complexity of \textit{LPT} is $O(n\log n)$, with an approximation ratio \cite{graham1966bounds} of $ZT^{max} = \left(\frac{4}{3} - \frac{1}{3z_l}\right)ZT^{OPT}$. Furthermore, we can also prove an upper bound for $\max_{j \le z_l}\{ZT_{l,j}\}$, as shown in Theorem \ref{theo:ZT_up}.


\vspace{-2mm}
\begin{theorem}
\label{theo:ZT_up}
Let $T_{l,v}^{rem} = \sum_{k=1}^{K_l}\left(PT_{l,k}^{rem} + \frac{2D}{B}N_{l,k}^{pre}\right)$, and $N^{up} = \frac{\sqrt{3N^{in}}}{2} + \frac{N^{in}}{K_l}$. Given $z_l$ replicas, With a high probability (95\%), $\max_{j \le z_l}\{ZT_{l,j}\} \le \frac{z_l-1}{z_l}[\sum_{v=1}^{V^e} y_{l,v}\tau_{l,v}^c(N^{up}) + \frac{2D}{B}N^{up}] + \frac{T_{l}^{rem}}{z_l} + t_{l}^{rem}$


\end{theorem}
\vspace{-1mm}

\subsubsection{\textbf{Remote Expert Replicas Decision}}

Theorem \ref{theo:ZT_up} provides the worst-case prefilling time $\max_{j \le z_l}\{ZT_{l,j}\}$, which enables us to optimize the replicas, $z_l$, to meet the TTFT constraint.

First, we initialize $Z=(z_1,...,z_L)$ to ensure each $z_l$ meets the payload size. Then, for each layer, we calculate the current replica potential:

\vspace{-5mm}
{\small
\begin{equation}
    \varpi(l, Z) = \{C^{loc} + C^{rem}\}_{Z,z'_l = z_l} - \{C^{loc} + C^{rem}\}_{ Z,z'_l = z_l + 1}
    \label{eq:replica_utility}
\end{equation}
}
\vspace{-5mm}


\noindent $\{C^{loc} + C^{rem}\}_{Z,z'_l = z_l + 1}$ represents the overall cost after $z_l$ increases by 1. For the layer with the greatest replica potential, $l^{max}$, we let the replicas of layer add one and update $Z$. This process is repeated until the worst-case TPOT is satisfied. Finally, if $\varpi(l,Z) > 0$ for some $l$, we continue to add replicas to reduce the overall cost until either $\varpi(l,Z) \le 0$ or $z_l = z^{\max}$ for all $l$.



%% file: evaluation.tex
\section{Evaluation}
\label{sec:evaluation}

\subsection{Settings}
\label{sec:settings}
\textbf{Testbed}. We implemented a prototype of \textit{Remoe} based on Kubernetes. It includes several key components: 1) To fit our inference framework, we modified all MoE models used in our experiments to support parallel inference with both local and remote experts. 2) We use the C++ LibTorch library and gRPC to provide efficient serverless inference services, minimizing data transfer overhead and response time. 3) Our Pod scheduler is NUMA-aware. The experimental platform is a server featuring a dual-socket configuration with two Intel Xeon Gold 6348 CPUs (totaling 56 cores, 112 threads). Furthermore, the server is equipped with two NVIDIA A100 GPUs, each providing 80 GB of VRAM.

\textbf{Dataset}. To ensure a comprehensive evaluation, our experiments are conducted on four widely-used datasets. These include: \textbf{LMSYS-Chat-1M} \cite{zheng2023lmsys}: A dataset with 1M real-world conversations for evaluating chat and instruction-following abilities. \textbf{WikiText-2} \cite{merity2016pointer}: A high-quality language modeling benchmark derived from Wikipedia articles. \textbf{C4} \cite{raffel2020exploring}: A massive, cleaned web-text corpus from Common Crawl, used for testing model generalization. \textbf{SlimPajama} \cite{cerebras2023slimpajama}: A large-scale and high-quality dataset designed for model pre-training.


\textbf{Models}. We use two MoE models at different scales: 1) GPT2-moe: The original GPT2 model has 12 hidden layers and 124 million parameters. The FFN of each layer is converted into 8 experts and a gating network. Each token is routed to 2 experts per layer for inference (remote expert memory specifications: [200, 2000] MB; main model: [200, 5000] MB). 2) Deepseek-v2-lite: It has 27 hidden layers and 16 billion parameters. Each layer has 64 experts and 2 shared experts except the first dense layer. Each token is routed to 6 experts and 2 shared experts per layer (remote experts: [1000, 5000] MB; main model: [1000, 40000] MB). The step size for memory specifications is 100 MB.

\subsection{Prediction Accuracy}

To evaluate the prediction performance of \textit{Remoe}, we compare it with the following baselines: \textbf{1) VarPAM.} Replace our customized k-medoids clustering with Partitioning Around Medoids algorithm \cite{kaufman2009finding}. \textbf{2) VarED.} Replace our distance metric during clustering, which is semantic similarity, with Euclidean distance between expert activation distribution matrices. \textbf{3) Distribution-Only Prediction (DOP) \cite{ma2025moe}.} Directly use the historical activation as prediction for new prompts. \textbf{4) Fate \cite{fang2025accurate}.} Predict expert activation per token using previous layer inputs. We adjust it by using the initial prompt embedding to predict activation across all layers for prompt-level prediction. \textbf{5) Equal Frequency (EF).} Assume that the activation frequencies of all experts are equal to each other. \textbf{6) Brute Force (BF).} Use brute-force searching to get top-$\alpha$ semantically similar historical prompts for the new one.

\begin{figure}[!htp]
\vspace{-5mm}
    \centering
    \subfloat[GPT2-moe]{\includegraphics[width=.48\columnwidth]{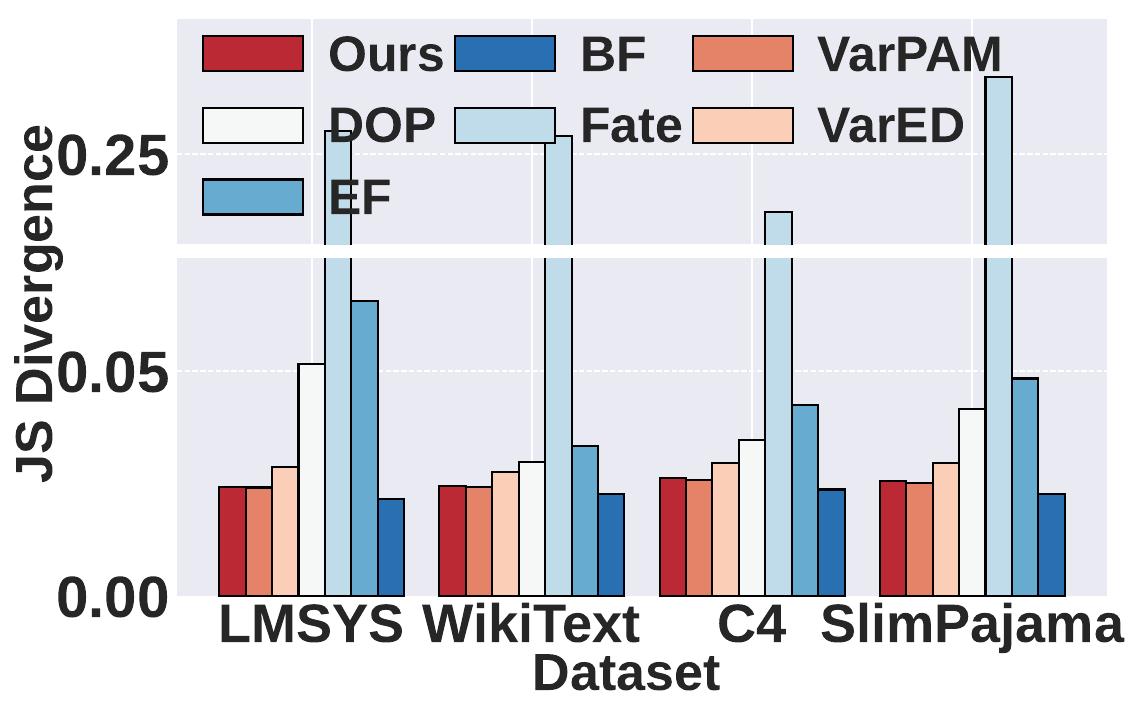}}\hspace{5pt}
    \subfloat[Deepseek-v2-lite]{\includegraphics[width=.48\columnwidth]{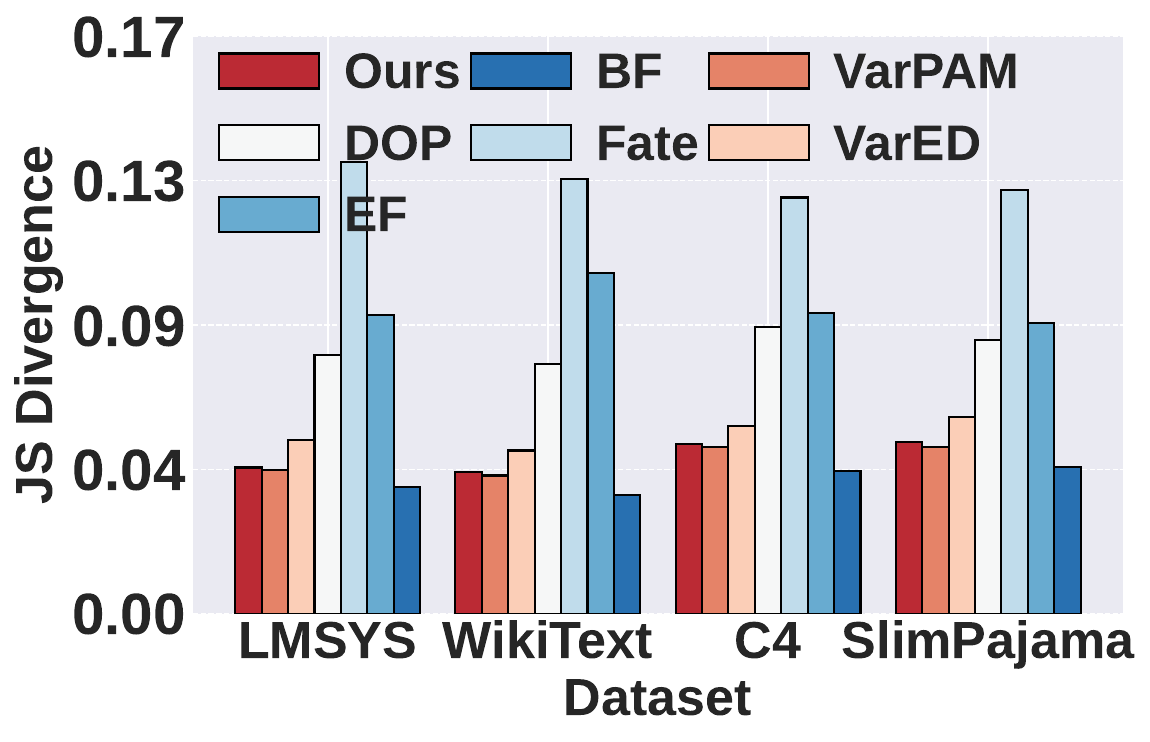}}\\
    \caption{JS Divergence under different datasets}
    \label{fig:predict_method_compare}
\vspace{-5mm}
\end{figure}

We randomly extract 5000 training and 500 test samples from each aforementioned dataset. Setting $\alpha=15$ and $\beta=150$, Fig. \ref{fig:predict_method_compare} shows our method achieves the lowest average JS Divergence (after VarPAM and BF) between predicted and true expert activation distributions. Partial truncation is applied to the y-axis in Fig. \ref{fig:predict_method_compare}a. Crucially, VarPAM requires hours to build its clustering tree (versus $\leq0.5$ seconds for ours), and our semantically similar prompts searching method is more than 10 times faster than BF. DOP is only effective when new prompt's expert activation is similar to historical ones, and Fate uses inappropriate inputs for prediction of various layers' expert activation. While expert activation and semantic similarity generally correlate, using expert activation distribution directly for clustering (VarED) introduces noise, explaining our superior performance.

\subsection{Overall Performance}

To evaluate the overall performance of \textit{Remoe}, we compare it with the following baselines: 
\textbf{1) CPU}. Deploy the MoE on CPU. \textbf{2) GPU}. Deploy the MoE on GPU. \textbf{3) Fetch}. The ideal situation for all expert offloading methods \cite{yu2025fmoe, song2024promoe, tang2024hobbit, xue2025moe}. It assumes that required experts are pre-loaded onto the GPU, with no mispredictions and no expert offloading/reloading time. \textbf{4)~MIX}. The expert modules are deployed on CPU, and other modules are deployed on GPU. The CPU and GPU memory are sufficient for modules caching.

\begin{figure}[!htp]
\vspace{-6mm}
    \centering
    \subfloat[GPT2-moe]{\includegraphics[width=.46\columnwidth]{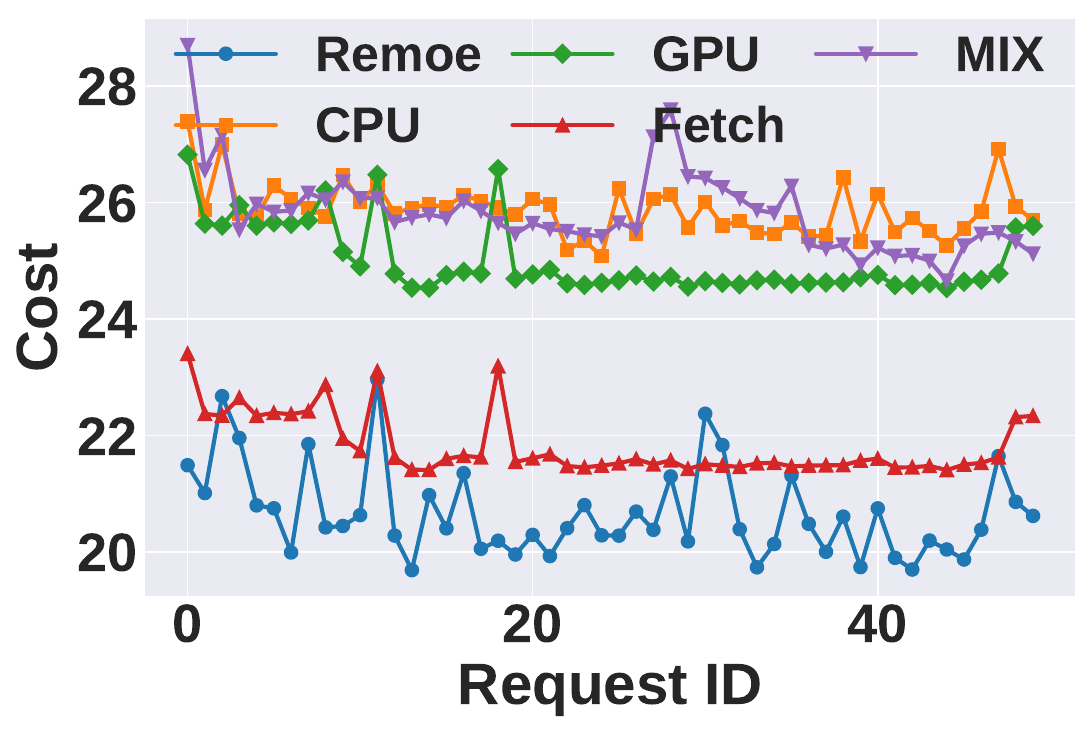}}\hspace{1pt}
    \subfloat[Deepseek-v2-lite]{\includegraphics[width=.48\columnwidth]{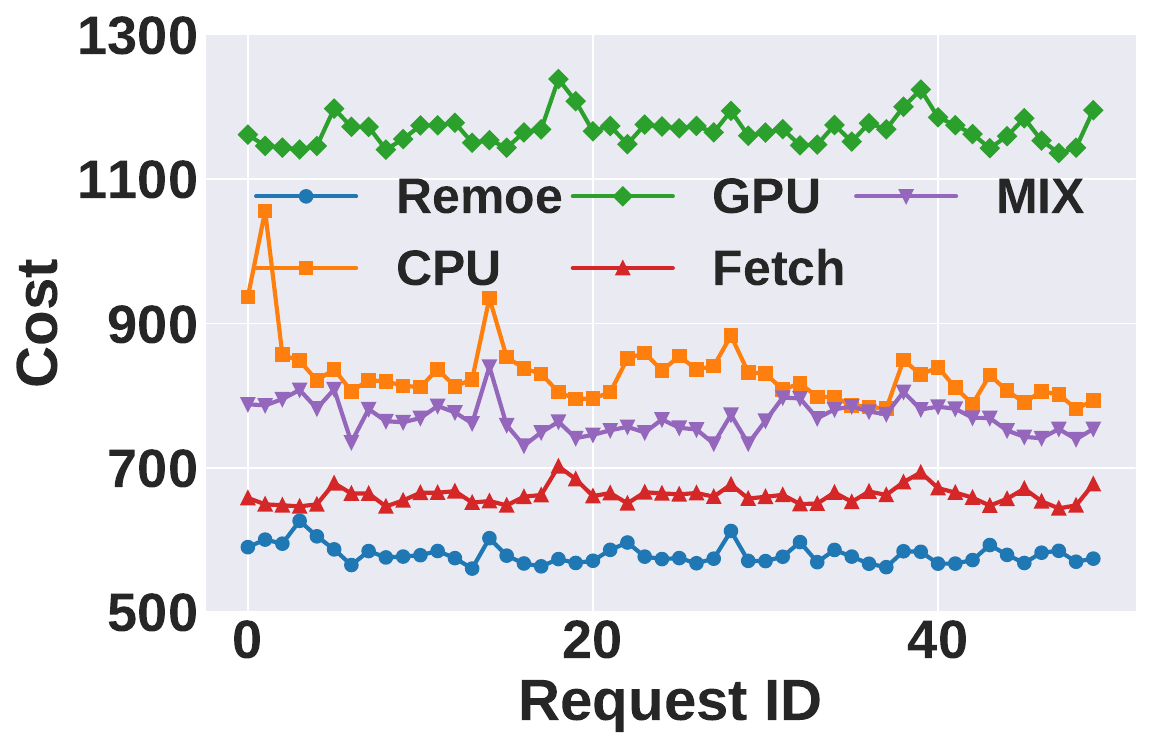}}\\[-3pt]
    \caption{Overall performance under 50 requests}
    \label{fig:overall}
\vspace{-4mm}
\end{figure}

We randomly sampled 50 tasks from the test set to serve as requests. For each request, we took the first 500 characters as the model input and set the number of output tokens to 200. Fig. \ref{fig:overall} shows the cost of the two models under different baselines. For both models, \textit{Remoe} achieves the lowest inference cost. It is observed that for the smaller MoE model (GPT2-moe), the cost difference among the methods is minor. For certain requests, Fetch even incurs lower cost than \textit{Remoe}. However, for the larger model (Deepseek-v2-lite), the cost differences become significant, with \textit{Remoe} achieving up to a 57.14\% cost reduction. Among these methods, MIX shows lower costs than GPU and CPU, demonstrating that a heterogeneous model can substantially decrease inference overhead. Meanwhile, although Fetch can theoretically achieve optimal performance, it still requires caching all experts in memory and needs additional GPU memory for loading partial experts. This characteristic introduces extra costs.

\subsection{Cost under Different Prefilling/Decoding Ratios}

\begin{figure}[!htp]
\vspace{-7mm}
    \centering
    \subfloat[GPT2-moe]{\includegraphics[width=.48\columnwidth]{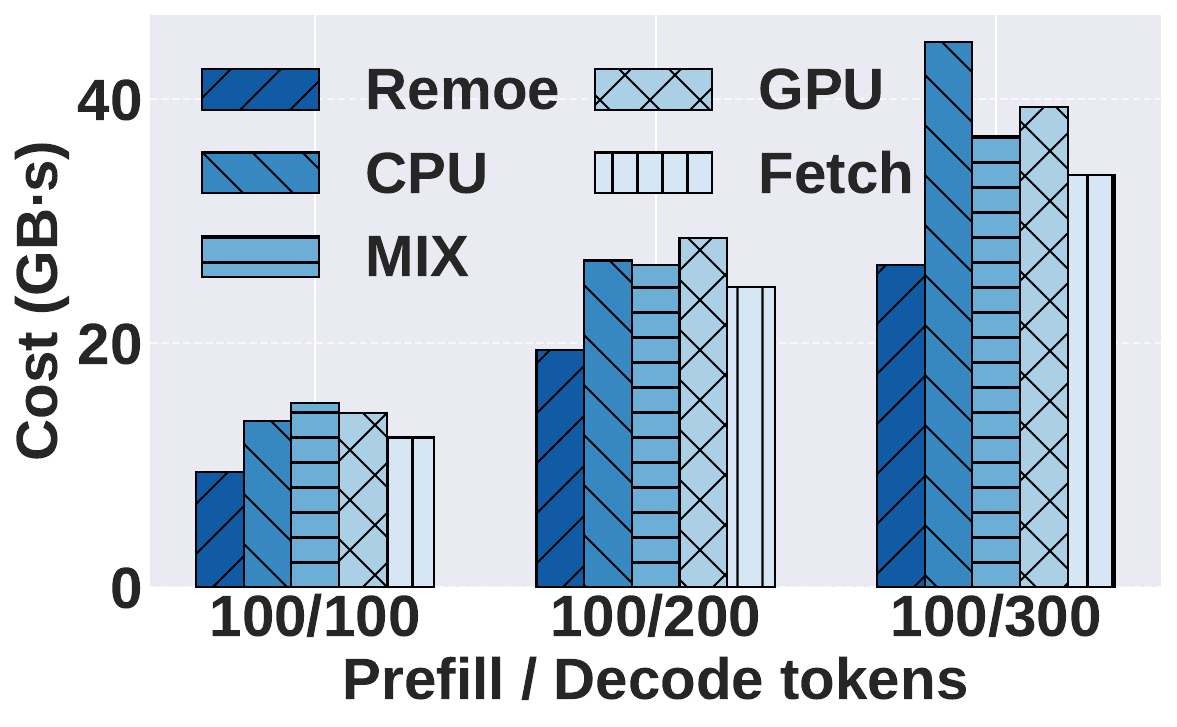}}\hspace{2pt}
    \subfloat[Deepseek-v2-lite]{\includegraphics[width=.48\columnwidth]{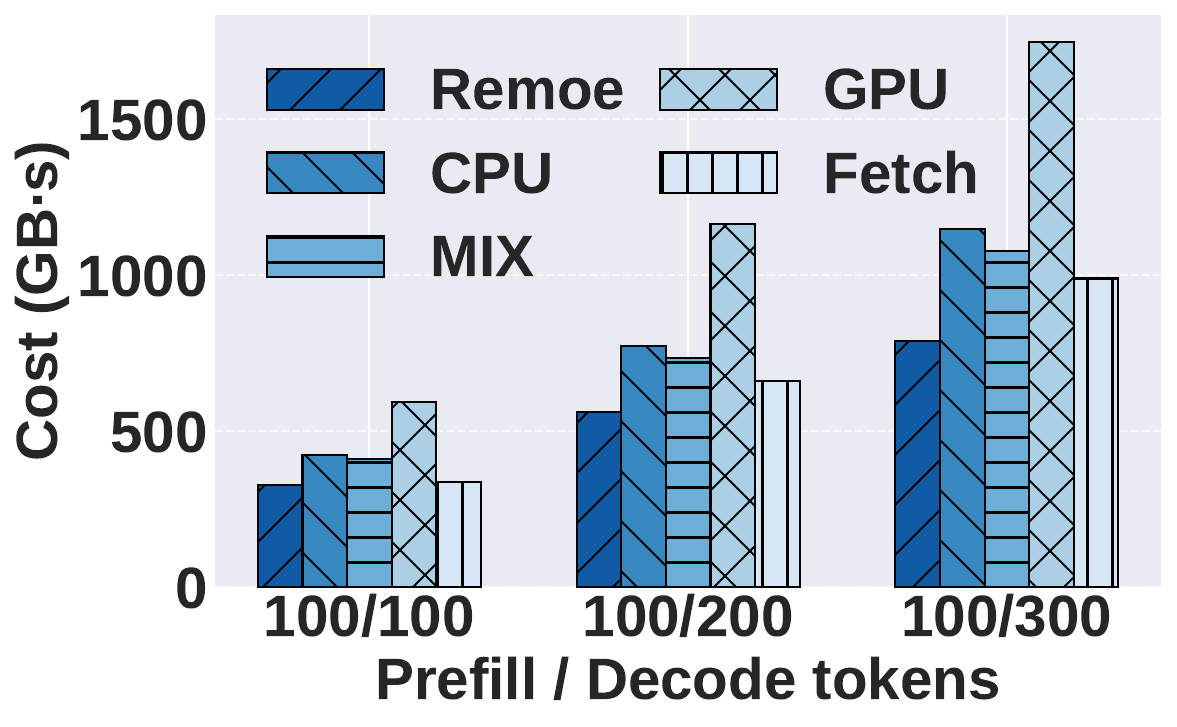}}\\[-4pt]
    \caption{Cost under different prefilling/decoding ratios}
    \label{fig:cost_tokens}
\vspace{-5mm}
\end{figure}

In real-world scenarios, the number of tokens in the decoding phase often exceeds that in the prefilling phase. Therefore, we study
the trend of inference cost under different ratios of prefilling to decoding tokens, as shown in Fig. \ref{fig:cost_tokens}. Across various ratios, \textit{Remoe} maintains stable performance. For GPT2-moe, as the number of decoding tokens increases, CPU's cost gradually surpasses that of other methods. Although deploying the model on CPU saves memory overhead, the longer inference time clearly negates this advantage. In contrast, for Deepseek-v2-lite, GPU's cost is significantly higher than other methods in all cases. This is because larger MoE models lead to more memory waste on low-frequency experts, especially for GPUs with higher pricing.

\subsection{Cold Start and Algorithm Overhead}


Cold start is a critical issue in serverless computing. As shown in Fig. \ref{fig:cost_tokens}, we compare the cold start times across different methods. While all approaches share the same container startup time due to a common base image, \textit{Remoe} achieves the lowest cold start time, with a reduction of up to 57.14\%. This improvement stems from its strategy of partitioning numerous experts into separate serverless functions, whose cold starts (labeled as REMOTE) can overlap with the main model's startup. Furthermore, Remoe's optimization logic (CALCULATE) is highly efficient; its overhead is negligible and introduces no additional waiting time.


%% file: related_work.tex
\section{Related Work}

\textbf{Serverless LLM Inference.}
Research on serverless LLM inference has focused on several key optimizations. To mitigate the cold start problem, techniques such as pipeline parallelism \cite{lou2025towards} and multi-tiered local storage \cite{fu2024serverlessllm} have been explored to accelerate model loading. Another key focus is resource allocation, where efforts include using elastic hardware sharing to boost GPU utilization \cite{xu2025llm} and combining adaptive configuration with real-time monitoring for stable serving \cite{huang2024enova}. For cost optimization, Liu et al. \cite{liu2025optimizing} proposed a specific scheduling algorithm ODS for serverless MoE inference, although limited to a pure CPU environment. Despite these advances, cost-efficient serverless MoE inference, particularly on GPU-CPU hybrid architectures, remains largely underexplored. 


\begin{figure}[!htp]
    \centering
    \subfloat[GPT2-moe]{\includegraphics[width=.9\columnwidth]{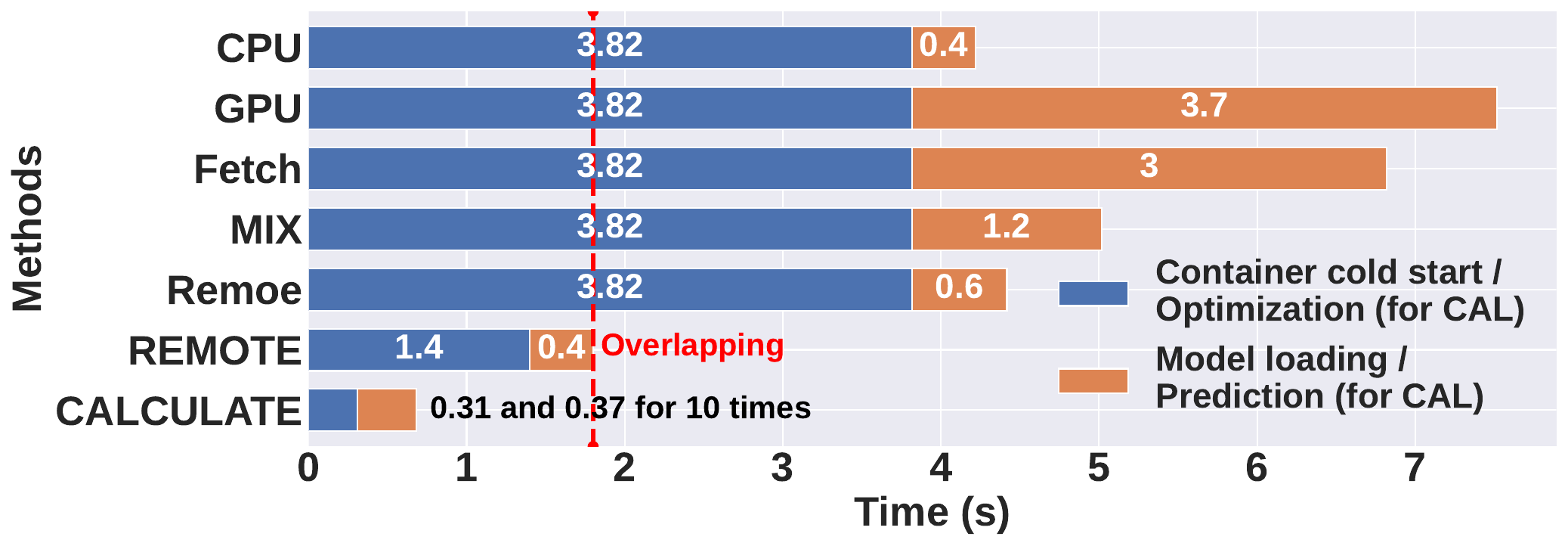}}\\
    \vspace{-4mm}
    \subfloat[Deepseek-v2-lite]{\includegraphics[width=.9\columnwidth]{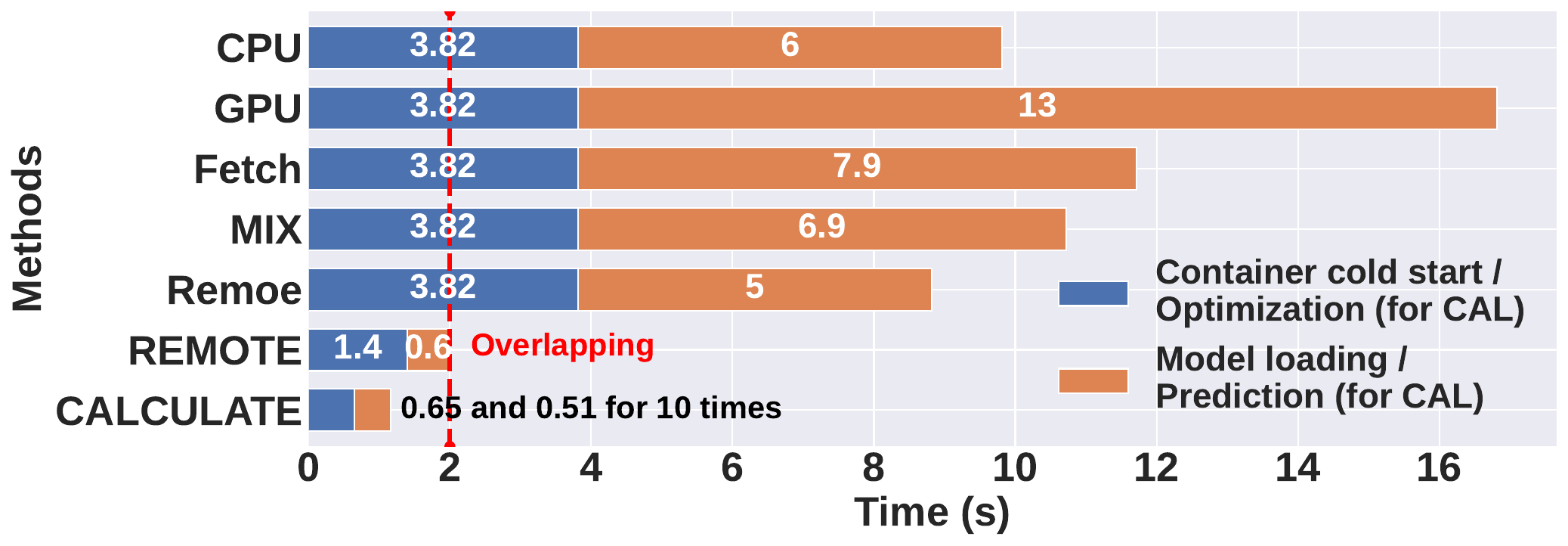}}\\
    \caption{Time for cold start, predicting, and optimization}
    \label{fig:cost_tokens}
    \vspace{-6mm}
\end{figure}


\textbf{GPU Memory-Constrained MoE Inference.} Prediction-based expert caching is the dominant approach for memory-efficient Mixture-of-Experts (MoE) inference. Strategies range from using historical data \cite{xue2025moe, yu2025fmoe} to more fine-grained, layer-level predictions, which have been successfully applied to memory-constrained devices with enhancements like mixed-precision loading \cite{tang2024hobbit} and graceful degradation \cite{zhang2025daop}. Another line of work employs dedicated ML predictors to achieve higher caching accuracy \cite{song2024promoe, he2024expertflow}. While effective, these token-level online prediction strategies are ill-suited for serverless environments that require resource pre-allocation, as the frequent adjustments would incur severe cold start overhead during execution.

%% file: conclusion.tex
\section{Concluding Remarks}
\label{sec:conclusion}

To minimize inference cost, we propose a heterogeneous system \textit{Remoe}. We design algorithms for expert activation prediction, resource pre-allocation, and joint memory-replica optimization. Our implementation of \textit{Remoe} on Kubernetes shows that it reduces inference cost and cold start latency significantly. Our current approach relies on idealized assumptions about the serverless environment. Consequently, our future work will focus on designing a highly fault-tolerant system to address real-world operational complexities such as unpredictable cold start times and network latency fluctuations.

%% file: appendix.tex
\appendix
\section{APPENDIX}
\label{appendix}

\subsection{Proof of Theorem \ref{theo:one_max_tokens}}

\begin{proof}
We prove it based on Hoeffding’s inequality. Due to space limit, all minor proofs (including Corollary \ref{cor:m_max_tokens}, Lemma \ref{lem:slater}, Theorem \ref{theo:opt_solu}, \ref{theo:ZT_up}) are deferred to the technical report~\cite{tech_report}.
\end{proof}

\subsection{Proof of Theorem \ref{theo:convex_analysis}}

\begin{proof}
First, taking the derivative of $g(\tilde{y}_l)$, we can obtain:

\vspace{-5mm}
{\small
\begin{align*}
    g'(\tilde{y}_l) = (c^c\theta_1-c^c\theta_1\theta_2\tilde{y}_l-H^w\theta_1\theta_2)\exp(-\theta_2\tilde{y}_l)+c^c(\theta_3+\frac{t_l}{\tilde{s}_l})
\end{align*}
}
\vspace{-5mm}

\noindent The second derivative is:

\vspace{-5mm}
{\small
\begin{align*}
    g''(\tilde{y}_l) = c^c\theta_1\theta_2^2\exp(-\theta_2\tilde{y}_l)[\tilde{y}_l-(\frac{2}{\theta_2}-\frac{H^w}{c^c})]
\end{align*}
}
\vspace{-5mm}

\noindent Since $c^c\theta_1\theta_2^2\exp(-\theta_2\tilde{y}_l) > 0$, $g''(\tilde{y}_l)$ is a monotonically increasing function. Its zero point is $\tilde{y}_l = \frac{2}{\theta_2}-\frac{H^w}{c^c}$. Therefore, when $\tilde{y}_l \ge \frac{2}{\theta_2}-\frac{H^w}{c^c}$, $g''(\tilde{y}_l) \ge 0$, and the function is convex.

Meanwhile, since $\frac{2}{\theta_2}$ is convex on $(0, \infty)$ and $\frac{H^w}{c^c}$ is constant, the function $\frac{2}{\theta_2} - \frac{H^w}{c^c}$ is also convex on this interval. Furthermore, since $\frac{d}{d\theta_2}\left(\frac{2}{\theta_2} - \frac{H^w}{c^c}\right) = -\frac{2}{\theta_2^2} < 0$, the term $\frac{2}{\theta_2} - \frac{H^w}{c^c}$ is monotonically decreasing in its domain. Therefore, when $\theta_2 \ge \frac{2c^c}{H^w}$, it implies that $\frac{2}{\theta_2} - \frac{H^w}{c^c} \le 0 < \tilde{y}_l$, ensuring $g(\tilde{y}_l)$ is strictly convex on $(0, \infty)$.
\end{proof}